\documentclass[aps,twocolumn,twoside,superscriptaddress,nofootinbib,floatfix]{revtex4-1}
\usepackage{amsmath,amsthm,amssymb,amsfonts}
\usepackage{graphicx}
\usepackage{siunitx}
\usepackage[pdftex]{hyperref}
\usepackage{xcolor}

\usepackage[T1]{fontenc}

\usepackage{tabularx}
\usepackage{booktabs}
\usepackage{multirow}
\newcolumntype{y}{>{\centering\arraybackslash}X}

\newtheorem{thm}{Theorem}

\newcommand{\appendices}{\appendix}

\newsavebox\figbox
\def\Figure[#1](#2)[#3]#4#5{%
  \setbox\figbox\hbox{\includegraphics[#3]{#4}}%
  \xdef\xfigwd{\the\wd\figbox}%
  \ifdim\wd\figbox<\columnwidth% Columnwidth Figure
  \begin{figure}[#1]%
    \centering%
    \includegraphics[#3]{#4}%
    \caption{#5}%
  \end{figure}
  \else% Wide Figure
  \begin{figure*}[#1]%
    \centering%
    \includegraphics[#3]{#4}%
    \caption{#5}%
  \end{figure*}
  \fi%
}

\begin{document}

\title{Timing constraints imposed by classical digital control systems on photonic implementations of measurement-based quantum computing}
\author{John R. Scott}
\affiliation{Quantum Engineering Centre for Doctoral Training, Department of Physics, University of Bristol, BS8 1FD, UK}
\author{Krishna C. Balram}
\affiliation{Quantum Engineering Technology Labs and Department of Electrical and Electronic Engineering, University of Bristol, BS8 1UB, UK}

\begin{abstract}
Most of the architectural research on photonic implementations of measurement-based quantum computing (MBQC) has focused on the quantum resources involved in the problem with the implicit assumption that these will provide the main constraints on system scaling. However, the `flying-qubit' architecture of photonic MBQC requires specific timing constraints that need to be met by the classical control system. This classical control includes, for example: the amplification of the signals from single-photon detectors to voltage levels compatible with digital systems; the implementation of a control system which converts measurement outcomes into basis settings for measuring subsequent cluster qubits, in accordance with the quantum algorithm being implemented; and the digital-to-analog converter (DAC) and amplifier systems required to set these measurement bases using a fast phase modulator. In this paper, we analyze the digital system needed to implement arbitrary one-qubit rotations and controlled-NOT (CNOT) gates in discrete-variable photonic MBQC, in the presence of an ideal cluster state generator, with the main aim of understanding the timing constraints imposed by the digital logic on the analog system and quantum hardware. We use static timing analysis of a Xilinx FPGA (7 series) to provide a practical upper bound on the speed at which the adaptive measurement processing can be performed, in turn constraining the photonic clock rate of the system. Our work points to the importance of co-designing the classical control system in tandem with the quantum system in order to meet the challenging specifications of a photonic quantum computer. 
\end{abstract}
 
\maketitle

%% Paper contents starts here %%%%%%%%%%%%%%%%%%%%%%%%%%%%%%%%%%%%%%%%%%%%%%%

\section{Introduction}
\label{sec:introduction}
Quantum computers are enjoying a period of intense research activity. This is due to the possibility of very large speed-ups in finding effective solutions for certain classes of problems which are hard or impossible to solve using classical computers. These include, in the near-term, the simulation of other quantum mechanical systems with applications in quantum chemistry \cite{2101.08448}, and in the longer term certain kinds of search-related problems \cite{Montanaro2016}. 

Currently, only relatively small quantum computers have been built, containing less than one hundred qubits, in a multitude of competing technologies \cite{Gyongyosi2019}. However, it is hoped that with increasing understanding over how to engineer and control quantum systems at scale, there will be a substantial increase in the computational power of quantum computers in the near future.

However, a few hurdles lie in the way of this scaling process. Often discussed are the characteristics of the qubits themselves: for example, the difficulty of achieving high enough fidelity gate operations on physical qubits \cite{Arute2019}, or how to fabricate devices at scale \cite{Osman2021}. One aspect of the problem, which is discussed less often, is the classical electronic control requirements that need to be met for for these machines.

At first glance, this is surprising since the common link between all the platforms for quantum computing is the need for traditional electronics to control them. However, there is a general tendency to view classical signal processing as a `solved problem', at least in comparison to the difficultly of the quantum information processing, and assume that any requisite performance can be achieved using custom application-specific integrated circuits (ASICs). This is understandable, given the high-degree of performance and sophistication that modern (digital) CMOS electronics can routinely achieve. However, adapting and applying classical electronics to quantum control problems is not straightforward and is an active area of research across all quantum computing platforms. For example, in superconducting qubit based quantum computers, it is an open question how best to integrate the microwave control electronics close to the qubits in order to reduce the number of interfacing wires that prevent scalability in that architecture. Novel optical routes for microwave delivery are being investigated as a potential solution to this problem \cite{Lecocq2021}.

In addition, the flip-side to modern CMOS electronics having reached the great level of sophistication that it enjoys today is that there is not a great deal of room left for improvement in performance. For example, clock speeds in computers are plateauing \cite{Ross2008}, and transistor sizes are reaching their limits \cite{Theis2017}. Quantum devices, on the other hand, being a relatively young technology, are expected to see a Moore's law-like improvement in the future \cite{Arute2019}. Consequently, it is critical to understand the limitations that classical electronics and control will impose on current quantum computing platforms.

The majority of the prior work discussing photonic MBQC has focused on the quantum resources required and the theoretical architecture of the system ~\cite{Browne2005,GimenoSegovia2015,Rudolph2017,2101.09310,Bourassa2021}. To our knowledge, the question of how far these schemes can be successfully implemented using current electronic devices remains unaddressed. As a first step towards this problem, we analyse the timing constraints imposed by classical electronics on the operation of a photonic quantum computer based on an ideal cluster state generator. We focus on photonic MBQC because of its relative insensitivity to photon loss in contrast to the gate based models. In addition, it is the primary architecture of choice for most active large-scale implementations of photonic quantum computers \cite{2101.09310,Bourassa2021}.

There are several important constraints the electronic control system may impose on the implementation of photonic MBQC. Firstly, the speed at which the electronics can be made to operate determines the maximum photon clock cycle of the system. In particular, the `flying-qubit' architecture of photonic MBQC (discussed in Section \ref{sec:simple-model} below) requires adaptive measurement settings to be worked out before the arrival of the next column of photons in the cluster state. Secondly, the complexity of the system -- particularly the analog parts -- determines how much on-chip area is taken up with classical processing. This is particularly important because the ``unit cell'' of the electronic system must be duplicated once per logical qubit in photonic MBQC, and signal routing becomes challenging as the system starts to scale. Thirdly, the noise that is introduced by the analog stages of the control system will potentially introduce logical errors that need to be accounted for.

In this work, we focus on the first of these questions: specifically, what timing constraints does the digital part of the system impose upon the analog and wider photonic systems, and how does that affect the overall photonic clock rate of the quantum computer? We perform the analysis for an idealized system with an ideal cluster state generator and consider the signal delay in the digital domain (after photon detection and logic-level amplification to the input of the analog system which is needed to set the bases for the next measurement round; details in Figure \ref{fig:analog-digital} below).

The structure of this paper is as follows. In Section \ref{sec:mbqc} we provide a practical description of the parts of MBQC needed to understand the results of this paper. In Section \ref{sec:simple-model}, we describe a simple model of photonic MBQC -- based on an ideal cluster state source -- showing the analog and digital systems which are necessary for a basic implementation of the system. In Section \ref{sec:fpga-design}, we present an example design for the digital component of the classical processing, targeting a Xilinx 7-series FPGA. In Section \ref{sec:verification}, we describe the functional verification of the design. In Section \ref{sec:timing}, we use static timing analysis of the implemented design to derive constraints on the analog parts of the system, and on the overall photonic clock frequency of the quantum computer. Section \ref{sec:extensions} contains changes to the underlying model to make it more realistic, which would increase the complexity of the digital system. Finally, we discuss the wider implications of our results in Sections \ref{sec:discussion} and \ref{sec:conclusion}.

\section{Introduction to Measurement-based quantum computing}
\label{sec:mbqc}

Quantum computing in the gate-based model (see Appendix \ref{app:gate-based-qc} for details) consists of the following steps:

\begin{enumerate}
\item An initial quantum state $|\phi\rangle$ is prepared on $N$ qubits;
\item Quantum gates are applied to the qubits;
\item The resulting state $|\psi\rangle$ is measured, which constitutes the output from the quantum circuit.
\end{enumerate}

MBQC is a different way to obtain the same resulting output state $|\psi\rangle$, by performing single-qubit measurements on a more complicated initial state called a cluster state. It consists of the following steps:

\begin{enumerate}
\item Prepare a special quantum state, called a cluster state, on a larger number $M>N$ of qubits. The main feature of the cluster state is that adjacent qubits are entangled together, which is represented using line segments in Figure \ref{fig:cluster-state}a;
\item Measure qubits from the cluster state one at a time, according to rules that correspond to the quantum circuit, until all but $N$ have been measured;
\item  Finally, the resulting state $|\psi^\prime\rangle$ on the $N$ remaining qubits is measured in the computational basis, which constitutes the output from the circuit. 

\end{enumerate}

The initial state $|\phi\rangle$ in the gate-based model is a matter of convention; if each qubit is initially prepared in the $|+\rangle$ state, then the output states $|\psi\rangle$ and $|\psi^\prime\rangle$ from the gate-based model and MBQC are the same, meaning that any algorithm expressed in the gate-based model can be equally well performed using MBQC.

For a comprehensive overview of MBQC, see \cite{Raussendorf2003}. A short pedagogical introduction is contained in \cite{quant-ph/0603226}. What follows is a brief description of the main features of MBQC which are relevant to this paper.

\Figure[t!](topskip=0pt, botskip=0pt, midskip=0pt)[width=0.9\columnwidth]{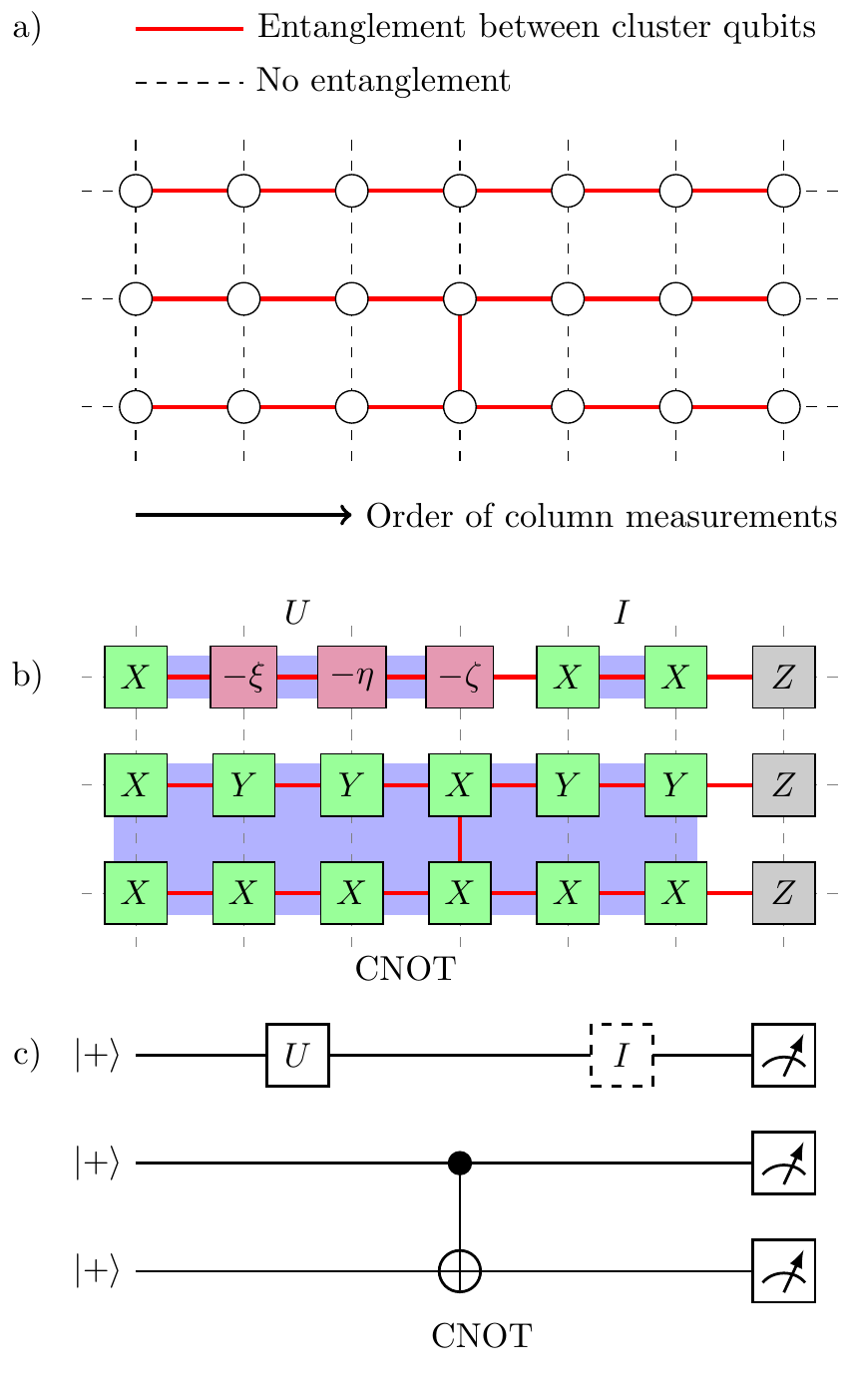}{a) A cluster state is made from a rectangular array of qubits (the white dots), each of which may be entangled with its four nearest neighbours. When a computation is performed, a specific pattern of entanglement is required that matches the shape of the circuit. b) The quantum computation is performed  by measuring the cluster qubits in bases derived from the measurement pattern. The shaded blue regions show which cluster qubits are involved in implementing which gates. The identity gate is included to pad the length of the one-qubit gate $U=R_x(\zeta)R_z(\eta)R_x(\xi)$ so it matches the CNOT. c) The quantum circuit that is performed by the measurement pattern in b).\label{fig:cluster-state}}

\subsubsection{Logical qubits and gates in MBQC}

Each horizontal line of entanglement in the cluster state corresponds to a single qubit in the gate-based model, which we will call a logical qubit, to distinguish it from the cluster qubits that make up the cluster state. One-qubit gates in the gate-based model involve measurements of the cluster qubits along a logical qubit row according to rules that determine the basis settings of each measurement, and define what to do with the measurement outcomes. Two-qubit gates require vertical lines of entanglement which join the logical qubit rows together, as shown in Fig. \ref{fig:cluster-state}a between the second and third row. 

\subsubsection{Measurement patterns for gates}

Each gate $G$ that is implemented in MBQC is defined by a measurement pattern, which is a set of rules describing:
\begin{itemize}
\item How many cluster qubits are needed to realise the gate $G$ and what pattern of entanglement is necessary between those cluster qubits;
\item Which basis to use for each cluster qubit measurement;
\item How to process the outcomes from the cluster qubit measurements.  
\end{itemize}

A given computation involving multiple gates, such as the one shown in the gate-based model in Figure \ref{fig:cluster-state}c, can be performed using MBQC by concatenating\footnote{In \cite{Raussendorf2003}, measurement patterns are taken to include the ``output'' qubits, which is the first column of qubits directly to the right of the measurement pattern. In this scheme, measurement patterns must overlap (because the output qubit column is also the input qubit column for the next gate pattern). In this paper, we associate the output qubit with the next measurement pattern, so that patterns can be simply concatenated.} the measurement patterns for each gate (the blue shaded regions in Figure~\ref{fig:cluster-state}b). The resulting pattern contains one row for each qubit in the gate-based model (here, $N=3$), and a number of columns defined by the length of the concatenated measurement patterns (the total number of cluster qubits is $M=21$).

In making the measurements defined by the measurement patterns, each cluster qubit is removed one by one until only the rightmost column remains unmeasured. The final column of the cluster state is measured in the computational basis as shown in Figure \ref{fig:cluster-state}b, which represents the output from the quantum circuit.

The arbitrary one-qubit gate $U$ in Figure \ref{fig:cluster-state}c is realised using a measurement pattern of four cluster qubits in the top row of Figure \ref{fig:cluster-state}b, and the CNOT gate\footnote{The symbol for a CNOT gate shown in Figure~\ref{fig:cluster-state} is the same as a classical XOR applied to the target qubit. This is because the CNOT gate can be thought of as adding the value of the control qubit to the target qubit modulo 2. We make extensive use of the classical XOR operation in subsequent figures in this paper. For clarity, we state here that all instances of the XOR symbol in this paper -- apart from in Figure~\ref{fig:cluster-state} -- are classical XOR gates, not CNOT gates.} is realised using a measurement pattern of 12 cluster qubits spanning the bottom two rows of Figure \ref{fig:cluster-state}b (note the vertical entanglement link).

All measurements shown in green and purple boxes in the figure are performed along lines $L$ that lie in the equator of the Bloch sphere (Figure~\ref{fig:bloch-sphere}). Green boxes containing $X$ or $Y$ are measurements along the $x$- or $y$-axis, respectively. Purple boxes are measured along a line $L$ with an angle $\phi$ derived from the value in the box and measurement outcomes of other cluster qubits. The grey boxes represent computational basis measurements, which are made along the $z$-axis of the Bloch sphere.

\Figure[t!](topskip=0pt, botskip=0pt, midskip=0pt)[width=0.99\columnwidth]{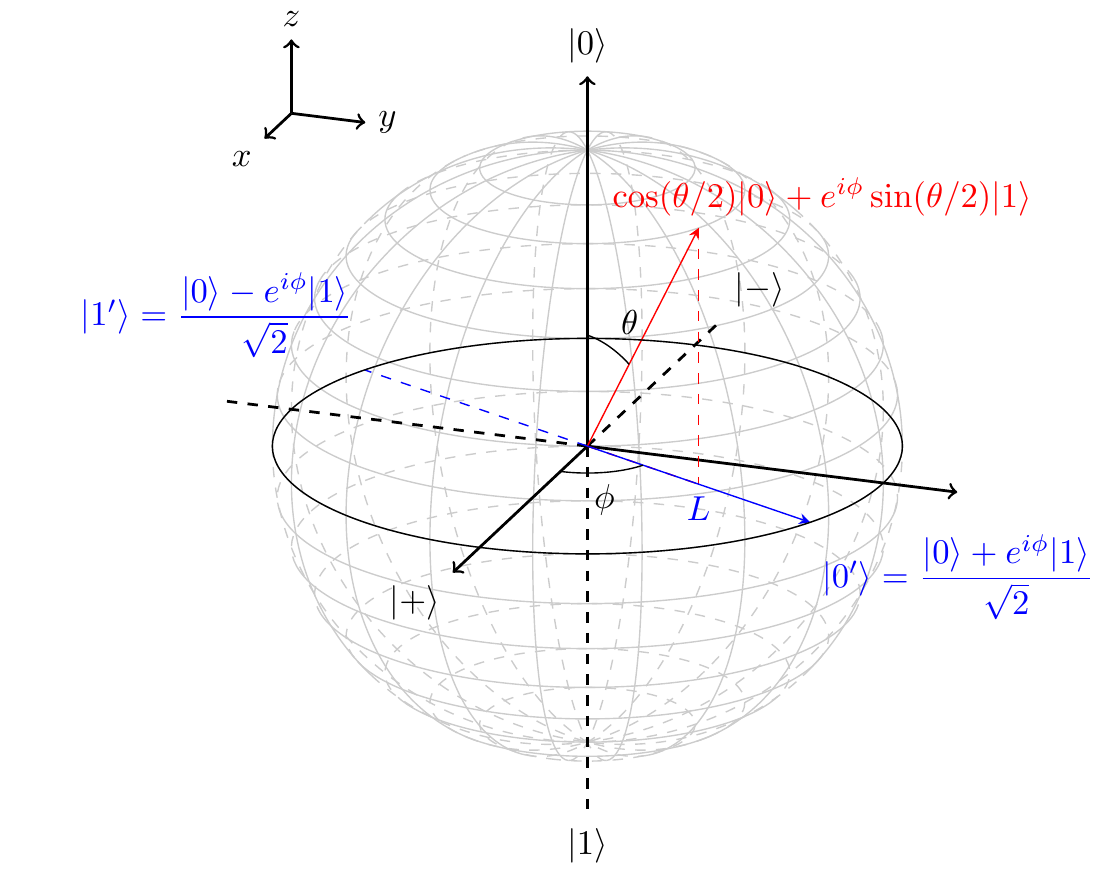}{The state of a single qubit can be represented as a point on the Bloch Sphere. A measurement of a single qubit can be made along any straight line through the Bloch sphere. In MBQC, measurements in the purple and green boxes in Figure \ref{fig:cluster-state} are made along lines $L$ in the equator of the Bloch sphere, parametrised by a single angle $\phi$. Computational basis measurements (denoted using grey boxes in Figure \ref{fig:cluster-state}) are made along the vertical line through $|0\rangle$ and $|1\rangle$.\label{fig:bloch-sphere}}

\subsubsection{Performing the cluster qubit measurements }

As described in Appendix \ref{app:gate-based-qc}, the only physical measurements that can be performed are computational basis measurements. All the other measurements (in the equator of the Bloch sphere) are performed by applying a one-qubit gate to the given cluster qubit and then measuring it in the computational basis.

It is important to understand that the one-qubit gates that set the measurement bases in the measurement patterns are different from the one-qubit gates implemented by MBQC, such as $U$ in Figure~\ref{fig:cluster-state}. The former are basic operations that, together with computational basis measurements, are required for implementation of MBQC. They are analogous to the physical layer in a communication system, because they must be realised by some physical mechanism; for example, using photonic qubits, as we discuss in Section~\ref{sec:simple-model}. The logical one-qubit gates $U$ do not correspond to any basic physical operation, and instead arise as a result of applying the measurement pattern to the cluster qubits. They are analogous to the logical data layers in a communication network, which use the resources of the physical layer to transmit logical information.

\Figure[t!](topskip=0pt, botskip=0pt, midskip=0pt)[width=0.99\textwidth]{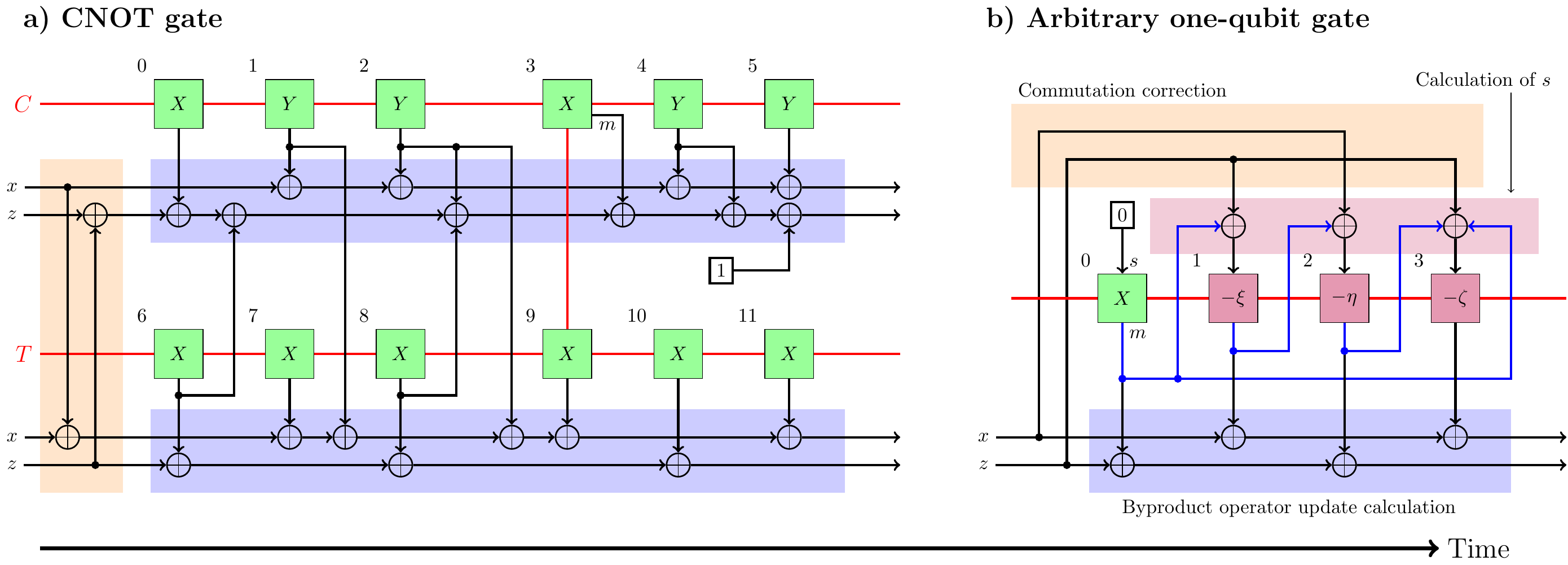}{The two measurement patterns we consider in this paper are the CNOT gate and the arbitrary one-qubit gate $U=R_x(\zeta)R_z(\eta)R_x(\xi)$. In each block, the black line connected to the top of the box is the adaptive measurement setting $s$. The line connected to the bottom of each box is the measurement outcome $m$. For the CNOT gate, on the left, there are no adaptive measurement settings, because all the measurement bases are $X$ or $Y$. However, the computation of the byproduct operators (shaded in blue) is more complicated, and involves mixing outcomes from the control $C$ and target $T$ rows. On the other hand, for the arbitrary one-qubit gate, the byproduct operator calculation is simple, but the adaptive measurement settings depend on previous measurement outcomes (shaded purple). The commutation correction for each gate is shaded in orange. For the CNOT gate, it involves mixing the byproduct operators before applying the pattern. For the one-qubit gate, the byproduct operators must be stored because they are used in the adaptive measurement setting calculation. In Section~\ref{sec:simple-model}, the condition is imposed that columns are measured from left to right, so as to be compatible with photonic MBQC.\label{fig:basic-patterns}}

In the following sections we describe in detail the measurement patterns for the one-qubit gate and the CNOT gate, which includes how to obtain the adaptive measurement settings and what to do measurement outcomes.

\subsubsection{Measurement basis angles and adaptive measurements}

Every measurement that is part of a measurement pattern is measured along a line $L$ in the equator of the Bloch sphere, as shown in Figure \ref{fig:bloch-sphere}. It is therefore specified by one real angle $\phi$. In MBQC measurement patterns, this angle is made up of a value $\theta$, and a sign bit $s$, such that $\phi = (-1)^s\theta$. The value of $\theta$ is shown in the purple boxes in Figure~\ref{fig:cluster-state}b. Note that $\theta$ may be negative.

For the green boxes in Figure~\ref{fig:cluster-state}, the value of $\theta$ is $0$ for $X$ and $\pi/2$ for $Y$. In the first case, the value of $s$ does not affect the basis angle $\phi$ at all. In the second case, the roles of $|0^\prime\rangle$ and $|1^\prime\rangle$ are swapped because $L$ is reversed; however, since the outcome of the measurement is random, the swapped measurement outcomes can be corrected in the calculation of byproduct operators (see Section \ref{sec:byproduct}). Therefore, the $X$ and $Y$ measurements are not affected by the value of $s$.

The value of $\theta$ is a characteristic of the quantum circuit being implemented. The value $s$, however, depends on the outcomes of other (prior) measurements in the measurement pattern. The measurement bases in MBQC are therefore adaptive, because the basis in which a cluster qubit is measured may depend on the outcomes of measurements of other cluster qubits which have been measured before. We will refer to $s$ henceforth as the adaptive measurement setting.

Any measurement pattern, such as the CNOT gate, which contains only $X$ and $Y$ measurements, does not involve adaptive basis settings, because the value of $s$ has no effect. It can be shown that the set of gates implementable with these non-adaptive patterns is the Clifford gate set \cite{Raussendorf2003}, which is not universal \cite{Nielsen2009}. For universal quantum computing, it is necessary to include a gate such as the one-qubit gate which does require adaptive measurement settings.

In Figure~\ref{fig:basic-patterns}b, the measurement pattern for the arbitrary one-qubit gate (corresponding to a rotation of the Bloch sphere) is shown in detail \cite{Raussendorf2003}. The shaded purple region (particularly the blue wires) shows how the adaptive measurement setting for each measurement is computed from previous measurement outcomes. The dependence between $s$ and measurement outcomes implies that the measurements must be made from left to right, which is also indicated by the arrow of time at the bottom of the figure.

The measurement pattern for the CNOT gate is shown in Figure~\ref{fig:basic-patterns}a. This is not the same pattern as that presented in the original MBQC paper \cite{Raussendorf2003}, which uses three logical qubit rows. The derivation of the CNOT pattern in Figure~\ref{fig:basic-patterns} is contained in Appendix \ref{app:cnot-derivation}. We use this modified CNOT measurement pattern because it considerably simplifies our example digital implementation in Section~\ref{sec:fpga-design}, which only supports nearest-neighbour connectivity of logical qubits.

\subsubsection{Byproduct calculations}
\label{sec:byproduct}

As the measurement pattern proceeds, the random outcomes of the measurements introduce correctable errors in the computation. These errors are known as byproduct operators, because they are unintended logical operations which occur as a byproduct of the MBQC measurements. 

Specifically, after any $N$-qubit gate $G$ has been applied to a state $|\psi\rangle$ using its measurement pattern, the resulting state is actually $BG|\psi\rangle$, rather than simply $G|\psi\rangle$, where $B$ is a gate (called the byproduct operator) given by
\begin{equation*}
  B = \prod_{i=1}^N Z_i^{z_i} X_i^{x_i},\qquad x_i,z_i\in\{0,1\}.
\end{equation*}

The byproduct operator for the logical qubit $i$ is specified by two bits $x_i$ and $z_i$, which are updated as the computation proceeds. By an abuse of notation, we will refer to the pair $(x_i,z_i)$ as the byproduct operator as well. For $N$ logical qubits ($N$ rows of the cluster state), $2N$ bits are needed to store the byproduct operators. At the start of the computation, they are all initialised to zero, because no gate has been performed so no errors have been introduced. As the computation proceeds, the outcomes of the measurements in the pattern are XORed into the $x_i$ and $z_i$ according to prescribed rules, described below and shown in Figure \ref{fig:basic-patterns}.

For the one-qubit gate in Figure~\ref{fig:basic-patterns}, the new byproduct operators $(x^\prime,z^\prime)$ are calculated according to the rule
\begin{align*}
  z^\prime &= z \oplus m_0 \oplus m_2\\
  x^\prime &= x \oplus m_1 \oplus m_3,
\end{align*}
where $m_k$ is the measurement outcome from the $k^\text{th}$ qubit, numbered according to Figure \ref{fig:basic-patterns}.

For the CNOT pattern, two byproduct operators are involved, one for the control qubit row $(x_c,z_c)$ and one for the target qubit row $(x_t,z_t)$. The new byproduct operators $(x_c^\prime,z_c^\prime)$ and $(x_t^\prime,z_t^\prime)$ are calculated using 
\begin{equation}
  \label{eq:cnot-byp-1}
  \begin{aligned}
  z_c^\prime &= z_c \oplus m_0 \oplus m_2 \oplus m_3 \oplus m_4 \oplus m_6 \oplus m_8 \oplus 1 \\   
  x_c^\prime &= x_c \oplus m_1 \oplus m_2 \oplus m_4 \oplus m_5,
\end{aligned}
\end{equation}
and
\begin{equation}
  \label{eq:cnot-byp-2}
  \begin{aligned}
  z_t^\prime &= z_t \oplus m_6 \oplus m_8 \oplus m_{10}\\  
  x_t^\prime &= x_t \oplus m_1 \oplus m_2 \oplus m_7 \oplus m_9 \oplus m_{11}.
\end{aligned}
\end{equation}

Unlike for the one-qubit gate, the byproduct operators for a given logical qubit row are calculated using measurements from other rows. Note the addition of the constant $1$ in the control qubit byproduct operator.

On the face of it, byproduct operators appear to introduce errors into the computation, because the gate $BG$ is performed instead of the desired gate $G$. However, the effect of this error can be corrected after the final column of $Z$-measurements in the MBQC process has been performed: the outcome from any logical qubit row $i$ where $x_i=1$ has its outcome flipped from a zero to a one or vice versa \cite{Raussendorf2003}. This action undoes the effect of the byproduct operators, leaving a circuit that effectively only implements the gate $G$ as desired. The $z_i$ components are not used because they correspond to a phase shift which does not affect the probability of measuring a zero or one in a computational basis measurement. However, as we describe in the next section, it is necessary to keep track of their values because they can affect the value of the $x_i$, through the process of commutation corrections.

\subsubsection{Commutation corrections}

The byproduct operators are used to correct the outcomes obtained after the MBQC circuit is finished. However, the correction only works if the byproduct operators are the last operation before the final column computational basis measurement, which is only the case if a single gate $G$ is performed.

If multiple gates $G_k$ are performed on a state $|\psi\rangle$, then the resulting state $|\phi\rangle$ will be
\begin{equation}
  |\phi\rangle = (B_KG_K)\dots(B_1G_1)(B_0G_0)|\psi\rangle.
\end{equation}
These interleaved byproduct operators cannot be corrected at the end of the circuit. Instead, it is necessary to move all the byproduct operators to the end (the leftmost side of the equation). To do that, after each new gate $G_{k+1}$ is applied, it is necessary to commute the current byproduct operators $B_k$ and the gate $G_{k+1}$, so that the byproduct operators are always on the leftmost side of the equation. This is illustrated below for the application of the second gate $G_1$:
\begin{multline}
  \label{eq:comm-correct}
  B_0G_0|\psi\rangle \mapsto B_1G_1B_0G_0|\psi\rangle\\\mapsto B_1B_0^\prime G_1^\prime G_0|\psi\rangle \mapsto B_r G_1^\prime G_0,
\end{multline}
where $G_1B_0=B_0^\prime G_1^\prime$, and the prime indicates the change that may occur in either gate. The byproduct operators $B_1$ and $B_0^\prime$ can be combined into a resulting byproduct operator $B_r$ by adding together the values of $(x_i,z_i)$ bitwise modulo 2 for each operator. The state on the right of Equation (\ref{eq:comm-correct}) is therefore transformed to the same form of the state on the left, so that on the application of the next gate $G_2$, the process can be repeated and the byproduct operators are always kept on the left. We will call the process of commuting $B$ through $G$ a commutation correction.

In practical terms, the commutation correction is an operation that is performed before a gate is applied, by manipulating the current value of the byproduct operators and the upcoming gate so as to have the effect of Equation (\ref{eq:comm-correct}). For the two measurement patterns we consider in Figure~\ref{fig:basic-patterns}, the commutation corrections are quite simple. In the case of the CNOT gate $G=\text{CNOT}$, $G^\prime = G$, and only the byproduct operator $B$ changes to $B^\prime$, according to the rule
\begin{equation}
  \begin{aligned}
    z_c^\prime &= z_c \oplus z_t \\
    x_c^\prime &= x_c \\
    z_t^\prime &= z_t \\
    x_t^\prime &= x_t \oplus x_c.
  \end{aligned}
  \label{eq:cnot-comm-correct}
\end{equation}

For the one-qubit gate $G=U$, the byproduct operators remain the same, $B^\prime=B$, but the gate itself $G$ must be modified. The modification is made by using the values of the byproduct operators to affect the adaptive measurement settings, by XORing the byproduct operators with previous measurement outcomes to form the values of $s$ for each cluster qubit\cite{Raussendorf2003}, as shown in Figure~\ref{fig:basic-patterns}b. The calculation of the adaptive measurment settings $s_j$ for each cluster qubit $j$ is shown in the following equations

\begin{equation}
  \begin{aligned}
    s_0 &= 0\\
    s_1 &= m_0 \oplus z\\
    s_2 &= m_1 \oplus x\\
    s_3 &= m_0 \oplus m_2 \oplus z.
  \end{aligned}
\end{equation}

It is necessary to make a copy of the byproduct operators $(x,z)$ before measuring the cluster qubits, because otherwise they will be overwritten during the calculations described in the previous paragraph. For example, after the measurement of cluster qubit 1 in the arbitrary one-qubit gate in Figure~\ref{fig:basic-patterns}, both the $x$ and $z$ values have been updated by measurement outcomes from cluster qubits 0 and 1. However, the old values of $x$ and $z$ are necessary in the measurement settings for cluster qubits 2 and 3.

In addition to the timing constraints imposed by the calculation of the adaptive measurement settings, the need to track the byproduct operators and calculate commutation corrections leads to additional timing constraints on digital implementations of the system, because they must be tracked in real time, and may affect adaptive measurement settings.

\subsubsection{Cutting out the right measurement pattern from the cluster state}
\label{sec:cut-out}

In the measurement pattern for the CNOT gate in Figure \ref{fig:basic-patterns}a, there are missing links between some of the cluster qubits. However, a fully connected cluster state does not contain missing links\footnote{Some authors distinguish a fully connected cluster state from a partial cluster state, which an example of the more general `graph state', because it has vertices corresponding to cluster qubits and edges corresponding to entanglement links. In this terminology, a fully connected cluster state is a graph state whose graph is the fully connected 2D lattice. To simplify our discussion, we refer to all graph states of any entanglement pattern as cluster states.}. In order for the CNOT measurement pattern to work, it is necessary to use an ideal cluster state generator, meaning one which can produce arbitrary patterns of nearest-neighbour entanglement in the cluster state.

In the more general approach to MBQC \cite{Raussendorf2003}, the computation always begins from the full cluster state. The links around a cluster qubit can then be removed by performing a computational basis measurement on that qubit. Using this method, cluster states containing less entanglement can be obtained from fully connected cluster states\footnote{This is one reason for using the three-row CNOT pattern, as in \cite{Raussendorf2003}: it is necessary to preserve a buffer row of cluster qubits between any two logical qubit rows that should not be connected. These buffer qubits are measured so as to remove the links between logical qubit rows.}.

The computational basis measurement that cuts out a cluster qubit incurs an additional step in the calculation of basis angles for surrounding cluster qubits. If the outcome of this measurement is a one, then a rotation $R_z(\pi)$ must be applied to the surrounding qubits, before they are measured according to any measurement pattern \cite{Raussendorf2003}. This gives rise to a more general form for the basis angle
\begin{equation}
  \label{eq:general-phi}
  \phi = \pi c + (-1)^s\theta,
\end{equation}
where $c$ is a single bit that is formed by XORing the measurement outcomes from any cluster adjacent qubits which have been removed using computational basis measurements.

Since the cut-out correction caused by removing a given cluster qubit must be performed before making the measurement of that qubit, the cutting out of cluster qubits introduces a measurement dependency between the measurement outcome of the cut-out qubit and the measurement angle $\phi$ of the qubits above and below it in the same column. This is different from our previous discussion on adaptive settings where the measurement outcomes had dependencies across columns in the cluster state and there was no intra-column dependence. This can only be handled by measuring each column in two rounds. First, the qubits that must be cut out are measured; then, the surrounding cluster qubits are measured using the modified basis angle $\phi$ in Equation (\ref{eq:general-phi}).

This aspect of cutting out the right shaped pattern from a fully connected cluster state can be avoided entirely if the right shaped cluster state is available from the beginning, using an ideal cluster state generator that can generate arbitrary patterns of entanglement. We assume the existence of such a cluster state generator for the purposes of this paper, and do not consider cut-out corrections any further.

\section{Simplified model of  photonic quantum computing}
\label{sec:simple-model}

In this section we describe how to implement MBQC using photons as qubits. We do not consider the generation of photonic cluster states, which is a separate subject in its own right~\cite{Browne2005,2101.09310}. Instead, we assume an ideal photonic cluster state generator, which can generate arbitrarily shaped rectangular cluster states, and describe how one can use it to perform photonic MBQC. We begin by describing how photons can be used as qubits.

\subsection{Quantum computing using photons as qubits}

In photonic quantum computing, a qubit is realised using a single photon. In the dual-rail encoding considered in this paper, a single photon passes through one waveguide or another depending on whether the qubit it represents is in the state $|0\rangle$ or $|1\rangle$, as shown in Figure \ref{fig:dual-rail-encoding}. A qubit encoded like this can be measured in the computational basis by placing a single-photon detector at the end of the pair of waveguides. It is important to realise that this process destroys the qubit (by absorbing the photon), unlike a matter-based qubit which can be re-used after measurement.

\Figure[t!](topskip=0pt, botskip=0pt, midskip=0pt)[width=0.9\columnwidth]{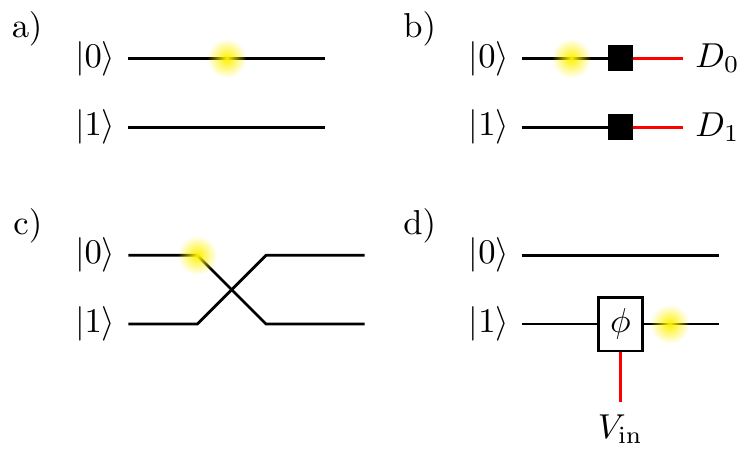}{a) A single photon in two waveguides can be used as a qubit. If the photon is in the top waveguide, then the qubit is in the $|0\rangle$ state, whereas if it is in the bottom waveguide, the qubit is in the $|1\rangle$ state. b) Computational basis measurements can be performed by placing a single-photon detector at the end of the waveguides. Basic one-qubit operations can be realised using linear optical elements such as c) beamsplitters and d) modulators. Complex operations can be realised by placing the elements one after the other.\label{fig:dual-rail-encoding}}

Modulators and beamsplitters can be used to realise an arbitrary one-qubit gate, as follows. First, a modulator in the $|1\rangle$ waveguide realises an arbitrary $z$-rotation, shown in Figure \ref{fig:dual-rail-encoding}d. Then, the variable beamsplitter shown in Figure \ref{fig:variable-beamsplitter} realises an arbitrary $x$-rotation. Finally, a second modulator in the $|1\rangle$ waveguide realises another arbitrary $z$-rotation, which completes the decomposition $R_z(\alpha)R_x(\beta)R_z(\gamma)$.

\Figure[t!](topskip=0pt, botskip=0pt, midskip=0pt)[width=0.65\columnwidth]{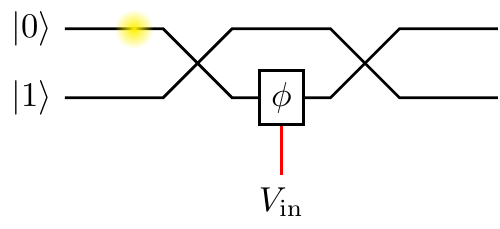}{A variable beamsplitter, which realises an $R_x(\phi)$ rotation, is formed by placing two fixed beamsplitters on either side of a modulator.\label{fig:variable-beamsplitter}}

In contrast to many other physical realisations of quantum computing, including superconducting qubits and trapped ions, that have a natural way to implement two-qubit operations~\cite{Krantz2019}, there is no simple deterministic way to implement the CNOT gate, or any other two qubit entangling gate, in terms of passive linear optical elements (modulators and beamsplitters). This is mainly due to the weakness of the direct photon-photon interaction. While this might appear to be a key limitation for photonic quatum computing, it was shown that one can implement an artificial non-linear gate that works probabilistically by using additional auxilliary photons and photodetection~\cite{Knill2001}. By parallel multiplexing these entangling gates, one can overcome their inherently probabilistic operation \cite{Rudolph2017}.

One of the arguments in favour of photonic MBQC is the absence of two-qubit gates in the implementation of a quantum circuit~\cite{Browne2005}; after the cluster state has been generated, only one-qubit gates and computational basis measurements are necessary. Much of the complexity is pushed to the task of generating the cluster state~\cite{GimenoSegovia2015}, which is responsible for all the entanglement between the qubits. As we describe in the next section, it is possible to generate the cluster state one column at a time, so that each photon only has to travel through a fixed length cluster state generating system, followed by a fixed length measurement system so that the overall photon loss can be bounded irrespective of the length of the equivalent quantum circuit (in the gate-based model) being implemented. Given photon loss is a primary source of error (and decoherence) for photonic quantum computing, this represents another important advantage of photonic MBQC~\cite{GimenoSegovia2015}.

\subsection{Photonic measurement-based quantum computing}

For matter-based implementations of MBQC, the grid of qubits directly corresponds to a two-dimensional physical array of atoms. However, for photonic quantum computing, it is not feasible to maintain a static array of qubits for long enough to perform the measurements. This is because a photon is always moving, so the only way to store it is to place it in a long waveguide, called a delay line, or keep it circulating in an on-chip cavity, such as a microring resonator. Both of these approaches eventually lead to photon decay, primarily due to scattering and absorption loss in the waveguide which is exacerbated in an integrated photonics platform (waveguide loss in a silicon platform is $\sim\SI{1}{\decibel\per\centi\meter}$ \cite{chrostowski2015silicon} compared to $\sim \SI{0.2}{\decibel\per\kilo\meter}$ for optical fibres \cite{Tamura2018}).

Instead, the cluster state can be generated one column at a time, and each column can be measured one after the other. This is opposite to the original presentation of MBQC \cite{Raussendorf2003}, where the goal was to separate the processes of generating the cluster state and making the measurements. The motivation for generating the cluster state all at once was also due to physical considerations: a matter-based cluster state can be generated using a tunable Ising interaction that acts globally on the system~\cite{quant-ph/0603226}. However, it can be shown in that the two approaches are equivalent \cite[Section II.D]{Raussendorf2003}; there, the successive column approach is used as tool for verifying measurement patterns.

When the cluster state generation and the photon measurement is alternated, a single photon only has to travel from its source, through the cluster state generator, through a fixed length waveguide, and finish at the measurement block.

For photonic MBQC, in Figure \ref{fig:basic-patterns}, the horizontal axis can therefore be interpreted as time, and the vertical axis as space. Each column of the cluster state is generated one at a time, progressing from left to right. Using this approach introduces a restriction which is not present in the matter-based realisation of MBQC. The scheme is only viable if the measurement settings for the currently measured block only depend on the outcomes of previously measured columns. This is quite a severe restriction, ruling out many of the measurement patterns originally proposed in \cite{Raussendorf2003} (for example the CPhase gate, a two-qubit gate that depends on a continuous parameter). However, this requirement is satisfied for the one-qubit gate and the CNOT gate described here. In the case of the CNOT gate, there are no measurement dependencies. For the one-qubit gate, all the measurement dependencies (the blue lines in Figure~\ref{fig:basic-patterns}b) point from left to right\footnote{In the context of photonic quantum computing, this is sometimes referred to as feedforward of measurement results.}.

\subsection{Timing constraints on the cluster state}

We do not consider the generation of the photonic cluster state, apart from making the following remark about the choice of time delay between the generation of columns, which is crucial for our timing analysis.

In order to entangle photons $P_n$ and $P_{n+1}$ from two adjacent columns $n$ and $n+1$ of the cluster state, they must be brought to the same location (for example, a beam splitter) at the same time. However, when performing the cluster qubit measurement for the MBQC measurement pattern, $P_n$ (from column $n$) must arrive at the detector a finite time before $P_{n+1}$ (from column $n+1$), to allow time for the processing of measurement settings, byproduct operators and commutation corrections. Therefore, $P_{n+1}$ must experience a delay $T_p$ (realised using an on-chip delay line or optical fibre) after the entangling operation of adjacent columns and the measurement block. The inverse of this delay $X_p = 1/T_p$ is the photonic clock frequency, which is the rate at which columns are produced and measured, and which determines the speed at which the quantum computation progresses.

Two distinct physical mechanisms provide upper and lower bounds for this delay. An upper bound is given by the loss of the on-chip delay line, optical fibre, or routing system involved in the delay of the photon. The lower bound is given by the time required to process the measurement outcomes. The object of our analysis is to estimate the lower bound. 

\subsection{The full MBQC system}

\Figure[t!](topskip=0pt, botskip=0pt, midskip=0pt)[width=0.99\textwidth]{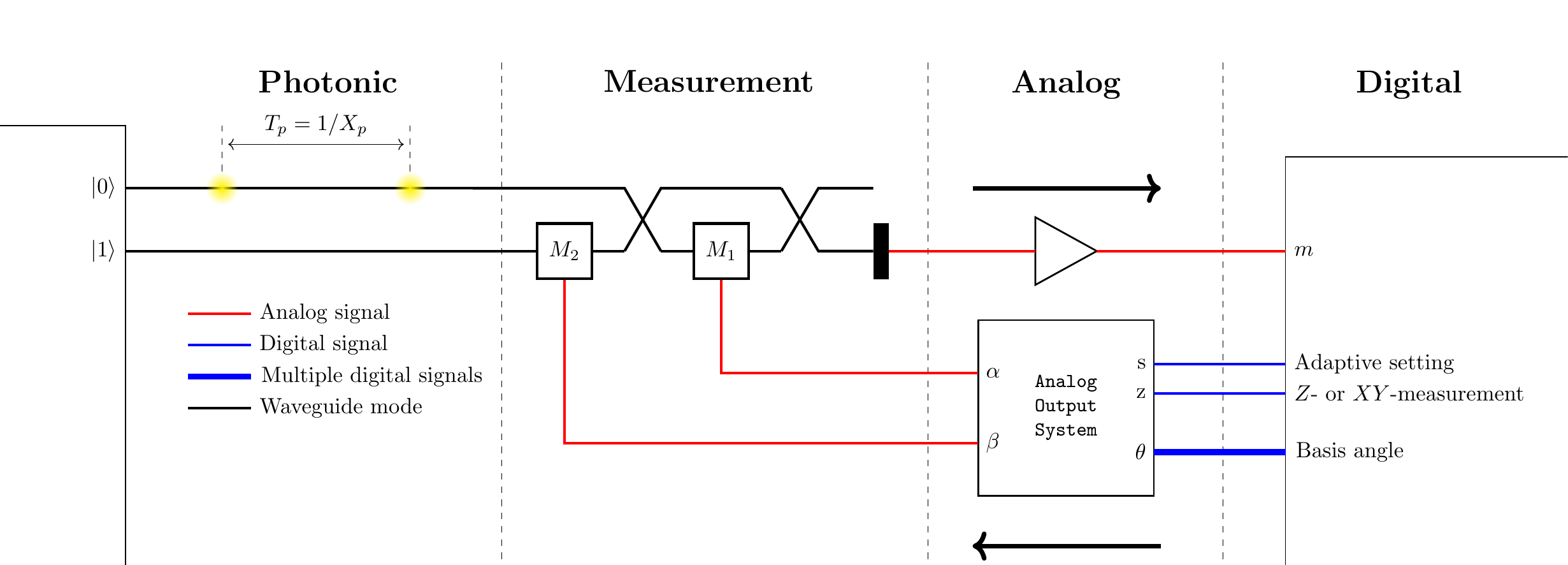}{One row of the system diagram of the classical control required to implement photonic MBQC. The cluster state generator is assumed to be ideal, outputting columns of photons at the photon clock frequency $X_p$. The cluster qubits represented by these photons are measured in bases specified by the measurement pattern in the measurement block, which is controlled by the voltages $\alpha$ and $\beta$ from the analog output system. The measurement results are amplified and processed by the digital system, which uses them to calculate subsequent adaptive measurement settings $s$ and byproduct operators. A copy of the system shown is required for each logical qubit, but each block is independent apart from the cluster state generator and the digital system.\label{fig:analog-digital}}

Figure \ref{fig:analog-digital} shows the full system required for processing one row of the MBQC measurement pattern, which corresponds to one logical qubit. It consists of the following six parts:

\begin{itemize}
\item The cluster state generator, which outputs the dual-rail encoded photon in each column of the cluster state one after the other. The photon has been entangled with the previous photon in the same row, and with the photons in the rows above and below as necessary for the measurement pattern.
\item The delay line, described in the previous section, which is necessary to temporally separate the photons in adjacent columns after they have been entangled.
\item The measurement block, which consists of passive linear optical elements that apply a configurable one-qubit operation, followed by a computational basis measurement.
\item The photon detector amplifier which converts the output from a single-photon detector to a logic level suitable for processing by a digital system.  
\item The digital system which processes measurement outcomes into adaptive measurement settings and keeps track of byproduct operators.
\item The analog output system, controlled by the digital system, which produces the analog voltage levels needed to drive the modulators in the measurement block.
\end{itemize}

The job of the digital system is to convert the measurement results into adaptive measurement settings for future measurements, and byproduct operators for interpreting the final measured outcomes.

The input to the digital system is the output pulse from the photon detector amplifier. This may be, for example, a superconducting-nanowire single-photon detector (SNSPD) \cite{Natarajan2012} followed by a low-noise amplifier \cite{Cahall2018}.

The output from the digital system includes the digital form of the angle $\theta$, the adaptive measurement setting output $s$, and a signal $z$ which determines whether the measurement is in the $XY$-plane of the Bloch sphere, or if it is a computational basis measurement.

\begin{table}[t]
  \centering
  \begin{tabularx}{\columnwidth}{ccc}
    \toprule
    z & Rotation & Measurement basis\\
    \midrule 
    \addlinespace[0.5em]
    1 & $\displaystyle R_z\left(\frac{\pi}{2}-\phi\right)R_x\left(\frac{\pi}{2}\right)$ & $XY$-measurement at an angle $\phi$\\
    \addlinespace[0.5em]
    0 & (None) & Computational basis measurement\\
    \addlinespace[0.5em]
    \bottomrule
  \end{tabularx}
  \caption{The table shows the one-qubit rotations generated by analog system modulator voltages. When $z=1$, a regular $XY$-basis measurement is performed, which accounts for the majority of cluster state measurements. A computational basis measurement is made at the end of the computation by setting $z=0$.}\label{tab:mod-voltages}
\end{table}

The analog output system is responsible for generating the voltages that control the modulators in the measurement block. It may be implemented using a combination of fast DACs and modulator drivers. Two modulators are necessary: one ($M_1$ in Figure~\ref{fig:analog-digital}) chooses between an $XY$-measurement and a computational basis measurement; and another ($M_2$) controls the basis angle $\phi$ for the $XY$-measurement. They are controlled by the voltages $\alpha$ and $\beta$ respectively, defined as follows\footnote{Voltages are expressed in modulator-phase units, where $V=1$ is chosen such that the modulator applies a \SI{1}{\radian} phase shift.}
\begin{equation}
  \label{eq:voltages}
  \begin{aligned}
    \alpha &= \frac{\pi}{2}z\\
    \beta &= \frac{\pi}{2}-\phi = \frac{\pi}{2}-(-1)^s\theta.\\
  \end{aligned}
\end{equation}

These modulator voltages realise the one-qubit rotation $R_x(\alpha)R_z(\beta)$, which sets the basis for the measurement. The one-qubit rotations are summarised in Table~\ref{tab:mod-voltages}.

The voltage $\alpha$ controls the $R_x$ rotation portion of the measurement setting, which determines whether the measurement is a computational basis measurement ($z=0$) or an $XY$-measurement ($z=1$). The voltage $\beta$ controls the angle of the $XY$-plane measurement $\phi$, which is itself determined by the fixed value $\theta$ and the adaptive measurement setting $s$.

In this paper, we focus on the digital control system, and present a simple reference design capable of performing the one-qubit gate and CNOT gate described in Section~\ref{sec:mbqc}. We analyse the timing behaviour of this design by implementing it with an FPGA and performing static timing analysis. The main objective of this analysis is to place timing constraints on the input and output analog systems, and therefore on the overall quantum photonic clock rate of the system. In the interest of simplicity, we ignore the final computational basis measurement of MBQC, which can easily be incorporated by setting $z=0$ for the final column of the pattern.

\section{Digital system design}
\label{sec:fpga-design}

In the following sections we describe an example digital system design\footnote{The design, along with other data used in the paper, is contained in the following repository: \url{https://gitlab.com/johnrscott/mbqc-fpga}.} for processing measurement outcomes into adaptive measurement settings and byproduct operators.

\subsection{Clock planning}

We present a design that can process measurements within a single clock cycle, by using three out-of-phase clocks. We consider a system synchronous design, with the photonic clock $X_p$ the common (master) clock in the system.

On the rising edge of $X_p$, the photon arrives in the measurement block, causing a pulse at the output of the single-photon detector. This measurement outcome is amplified and triggers a latch which provides a constant digital signal to the digital system.

The other two clocks, $X_s$ and $X_r$, are internal to the digital system. On the rising edge of the measurement sample clock $X_s$, the measurement latch is sampled by the digital system. The rising edge of $X_s$ must be sufficiently offset from the rising edge of $X_p$ so that the output from the latch has settled to a steady state. This delay must include the time required to amplify the photon detector output.

On the rising edge of the reset clock $X_r$, the latch is reset ready for the next measurement round. This event must occur after the rising edge of $X_s$, but before the rising edge of the next photon clock cycle $X_p$, to satisfy the hold time requirement of the sampling logic.

The computation of the adaptive measurement setting is performed using combinational logic at the earliest possible time that the latch output is valid, on the rising edge of $X_s$. The measurement setting for the next measurement is then computed and becomes available a short amount of time after the rising edge of $X_s$, corresponding to the combinational logic delay. 

In addition, the byproduct operators are also computed on the rising edge of $X_s$ using combinational logic. The commutation correction, which must be applied at the boundary of a quantum gate, is then computed on the rising edge of $X_r$, because it requires the value of the byproduct operators computed on $X_s$. The program which controls the measurement pattern is loaded from memory on $X_p$ so that it is ready for the computations that take place on $X_s$ and $X_r$.

The design of each computational subsystem is described in detail below. 

\subsection{Adaptive measurement setting generation}

\Figure[t!](topskip=0pt, botskip=0pt, midskip=0pt)[width=0.99\textwidth]{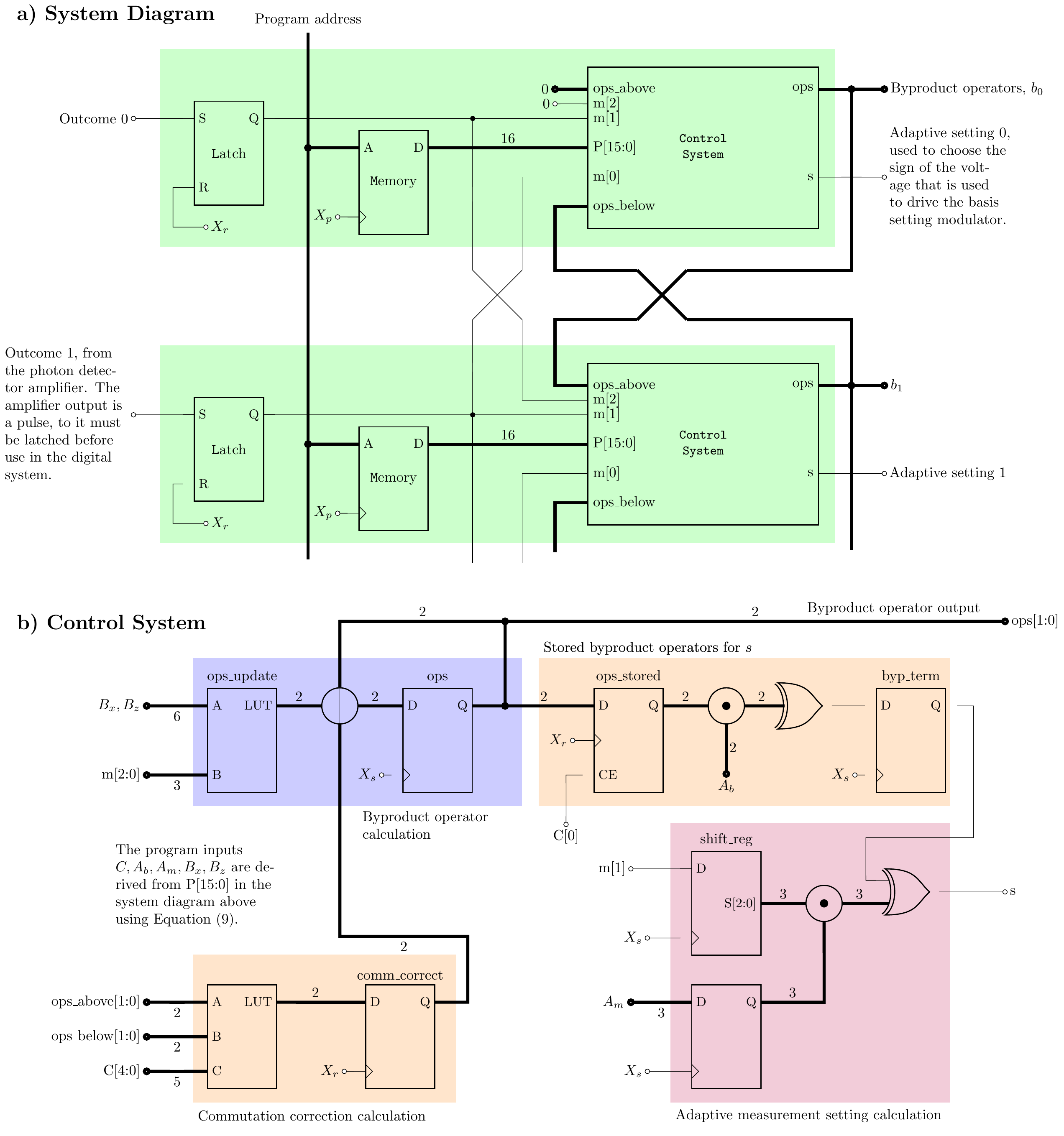}{a) The digital system diagram for multiple qubits. The ``unit cell'' for each qubit (shaded green) has a measurement latch, a program memory, and a control system for calculating measurement settings and byproduct operators. b) The design of the control system. In the high level schematic diagram of the control system, buses are denoted with bold lines, and the bus width is written next to the wire. The circles apply bitwise operations between their inputs: the cross stands for XOR and the dot stands for AND. The right port of the circle is the output, and all other ports are inputs. The logic gates are multi-input, with inputs from all the buses and wires connected on their left (i.e. wires inside a bus will be combined in the logic operation). Each part of the diagram is shaded according to its function, using the same colouring as in Figure \ref{fig:basic-patterns}. Flip-flops are clocked on the rising edge of their clock input, and elements whose output is \texttt{LUT} represent combinational logic. Reset signalling is omitted from the diagram for simplicity.\label{fig:system}}

The most important feature of the adaptive measurement setting $s$ is that it must be present as soon as possible, ready for the next measurement round. The earliest possible time that $s$ can be computed is on the rising edge of $X_s$. From Figure \ref{fig:basic-patterns}, the value of $s$ can depend on previous measurement settings and stored byproduct operator values from the current qubit.

A shift register is used to store the past three measurement values\footnote{For more complicated measurement patterns it may be necessary to store more than three measurements. However, for the arbitrary one-qubit gate and CNOT gate, three measurements are sufficient.}, $m_0$, $m_1$ and $m_2$, where $m_0$ is the most recent measurement outcome. The shift register is loaded sequentially with the next measurement on the rising edge of $X_s$. The output $s$ is then obtained using a combinational circuit from the shift register, so it is present soon after the rising edge of $X_s$.

The outputs from the shift register are combined bitwise with a 3-bit mask $A_m$ and XORed together to produce the measurement contribution to $s$. The stored byproduct operators $(x_s, z_s)$ are masked using a two-bit value $A_b$ and XORed to produce a second contribution to $s$. These two contributions are XORed to produce $s$ itself. Putting together these two contributions gives the following expression for $s$:
\begin{equation*}
  s = \left(\bigoplus_{i=0}^2 A_m[i]m_{i} \right)\oplus \left(A_b[1] x_s \oplus A_b[0] z_s\right),
\end{equation*}
where square brackets denote bitwise access.

The masks $A_m$ and $A_b$ for each measurement round are chosen in such a way that they combine past measurement outcomes and byproduct operators correctly to realise the one-qubit gate, as shown in Figure~\ref{fig:basic-patterns}b. The CNOT gate has no adaptive measurement settings, so $A_m = A_b = 0$ in that case.

The mask $A_m$ must remain valid through the rising edge of $X_p$, so it is registered on the rising edge of $X_s$. The byproduct operator contribution due to $A_b$ is also registered on $X_s$, so that the byproduct term persists through $X_p$. These registers are necessary because the program word, which contains the masks (see Section \ref{sec:program} below), is updated on the rising edge $X_p$.

A disadvantage of this design is that the output $s$ may contain function hazards \cite{Eichelberger1965}, due to the propagation delays from each of the flip-flops to the output $s$. These hazards do not affect the digital function of the (synchronous) digital system; however, they may contribute to the power dissipation of the system and/or noise in the analog output, depending how it is implemented. In order to avoid the hazards, the output $s$ could be registered; however, this would require another clock edge soon after $X_s$ to preserve the setup time of the analog output stage.

The adaptive system is shaded in purple in Figure~\ref{fig:system}b. 

\subsection{Byproduct operator calculation}

The byproduct operators must be updated after each measurement round. Since they only depend on the measurement outcomes, they can also be computed on the rising edge of $X_s$.

The byproduct operators comprise two bits $(x,z)$, which are updated according to the measurement outcomes from the current logical qubit, $m_0^{(1)}$, and the two neighbouring logical qubits, $m_0^{(2)}$ above and $m_0^{(0)}$ below. Any of these three measurements may be XORed in any combination, together with the old byproduct operator values $(x,z)$, to produce new $(x^\prime,z^\prime)$. Two 3-bit masks $B_x$ and $B_z$ control which of the three measurements outcomes should be XORed together to produce the updated $x$ and $z$, so that the byproduct operators are obtained using the following equations:
\begin{align*}
  x^\prime &= x \oplus \left(\bigoplus_{j=0}^2 B_x[j]m_0^{(j)}\right)\\
  z^\prime &= z \oplus \left(\bigoplus_{j=0}^2 B_z[j]m_0^{(j)}\right).
\end{align*}

The masks $B_x$ and $B_z$ for each measurement round are chosen in such a way that they combine measurement outcomes from the current and surrounding logical qubit rows to form the updates to the byproduct operators that are shown in Figure~\ref{fig:basic-patterns}.

It is sometimes necessary to add a constant (the 1 in Equation (\ref{eq:cnot-byp-1}) for $z_c^\prime$) to the byproduct operators, as in the case of the CNOT pattern. This constant addition is controlled by the commutation correction program, as described in the section below.

The main byproduct operator calculation is shaded in Figure~\ref{fig:system}b.

\subsection{Commutation corrections}

For the CNOT gate, the commutation correction is performed by mixing the values of the byproduct operators between the control and target logical qubits, as described in Equation (\ref{eq:cnot-comm-correct}).

For an arbitrary one-qubit gate, the correction is more complicated, requiring the use of the byproduct operators in the calculation of the measurement settings. However, in order to avoid overwriting these correctional byproduct operators prematurely, it is necessary to store them in a separate register, called the stored byproduct operator register. The correction for the one-qubit gate then amounts to loading this register from the current byproduct operators.

Both these corrections, for the CNOT and the one-qubit gate, require the byproduct operator values and must therefore be calculated on the rising edge of $X_r$ rather than $X_s$. The behaviour of this correction is controlled by a 5-bit value $C$, whose interpretation is shown in Table \ref{tab:c-prog}.

Most of the time ${C=0}$ and the commutation correction does nothing. It is only directly before gate boundaries that a commutation correction must be performed. 

The commutation corrections are shaded in orange in Figure \ref{fig:system}b.

\begin{table}[t]
  \centering
  \begin{tabularx}{\columnwidth}{l lX@{}}
    \toprule
    Bit & Meaning if high\\
    \midrule
    0 & Store the byproduct operators\\
    \midrule
    1 & Commutation correction is necessary, in which case:\\
    2 & \qquad If high then current logical qubit is the control\\
    3 & \qquad If high then other qubit in CNOT is above\\
    \midrule
    4 & Add ones to the byproduct operators, in which case:\\
    2 & \qquad Contains the constant value to add to z\\
    3 & \qquad Contains the constant value to add to x\\
    \bottomrule
  \end{tabularx}
  \caption{The table contains the interpretation of the bit fields of $C$, which controls the commutation correction for the arbitrary one-qubit gate and CNOT gate, and also controls the addition of constants to the byproduct operator. The meaning of bits 2 and 3 depend on whether bits 1 or 4 are set, which are mutually exclusive. If $C=0$ then no operation is performed. }\label{tab:c-prog}
  
\end{table}

\subsection{Program word}
\label{sec:program}

The digital system is controlled using a 16-bit program word $P$ which is formed by concatenating the masks and control bits in the previous sections as follows:
\begin{equation}
  P = CA_bA_mB_xB_z.
\end{equation}
Each logical qubit requires its own set of program words, one per measurement round. 

Table \ref{tab:program} shows an example calculation for the two qubit circuit containing an arbitrary one-qubit gate ${U=R_x(0.3)R_z(0.2)R_x(0.1)}$ on the first qubit, followed by a CNOT gate between the first and second qubit. The table contains randomly chosen measurement outcomes and the resulting adaptive measurement settings and byproduct operators that result from the measurement pattern, including the program word that is used to make the calculations. 

\begin{table}[t]
  \centering
  \begin{tabularx}{\columnwidth}{c *{2}{y} cc *{3}{y} cc *{2}{y}}
    \toprule
    && \multicolumn{5}{c}{Qubit 0} & \multicolumn{5}{c}{Qubit 1}\\
    \cmidrule(lr){3-7}\cmidrule(lr){8-12}
    Gate & $A$ & $m_0$ & $P_0$ & $\theta_0$ & $s_0$ & $b_0$ & $m_1$ & $P_1$ & $\theta_1$ & $s_1$ & $b_1$\\
    \midrule
    $U$ &0 & 0 & 0302 & 0 & 0 & 00 & 0 & 0002 & 0 & 0 & 00 \\
    &1 & 1 & 0510 & -0.1 & 1 & 10 & 1 & 0010 & 0 & 0 & 10 \\
    &2 & 1 & 0342 & -0.2 & 1 & 11 & 0 & 0002 & 0 & 0 & 10 \\
    &3 & 0 & 3010 & -0.3 & 0 & 11 & 1 & 5010 & 0 & 0 & 00 \\
    \midrule
    CNOT &4 & 1 & 0003 & 0 & 0 & 10 & 0 & 0002 & 0 & 0 & 10 \\
    &5 & 0 & 0010 & $\pi/2$& 0 & 10 & 1 & 0030 & 0 & 0 & 00 \\
    &6 & 0 & a013 & $\pi/2$ & 0 & 10 & 0 & 0022 & 0 & 0 & 00 \\
    &7 & 1 & 0002 & 0 & 0 & 10 & 1 & 0010 & 0 & 0 & 10 \\
    &8 & 1 & 0012 & $\pi/2$ & 0 & 01 & 0 & 0002 & 0 & 0 & 10 \\
    &9 & 1 & 0010 & $\pi/2$ & 0 & 11 & 0 & 0010 & 0 & 0 & 10 \\
    
    \bottomrule
  \end{tabularx}
  \caption{Example two-qubit computation comprising a one-qubit gate $U=R_x(0.3)R_z(0.2)R_x(0.1)$ on qubit 0, followed by a CNOT between qubits 0 and 1 (qubit 0 is the control). The program $P_i$ (written in hexadecimal in the table) combines the measurement outcomes $m_i$ (randomly generated) to produce the adaptive measurement setting $s_i$ and the byproduct operators $b_i$ (the least significant bit is $z$) for the $i^\text{th}$ qubit. The basis measurement angles $\theta_i$ are included for completeness ($s_i$  is combined with $\theta_i$ to produce the measurement angle $\phi_i$).}\label{tab:program}
  
\end{table}
  
It is clear that the program word could be compressed to save on memory usage. In our example design, we have prioritised program simplicity over memory usage.

\subsection{FPGA Implementation of the design}

In order to analyse the timing characteristics of the system, we wrote an FPGA implementation of the design using VHDl, targeting a Xilinx Kintex-7 FPGA (part no. xc7k70tfbg484-2). We used the synthesis tool Xilinx Vivado 2020.2 to implement the design and perform static timing analysis.

We used the mixed-mode clock manager (MMCM) \cite{ug472} to generate the two out-of-phase clocks $X_s$ and $X_r$ from the (external) system clock $X_p$. The program was stored in memory generated by an instance of the distributed memory generator IP \cite{pg063}, configured as ROM so that we could store the program in a coefficients file for the purpose of the verifying the design.

The utilisation of logic and input/output (I/O) pads in the design is provided for 1 logical qubit and 20 logical qubits in Table \ref{tab:util}. The data was obtained from the utilisation report generated by Vivado after implementing the system for each number of logical qubits. The number of logic elements scales more than linearly between 1 and 20 logical qubits because the synthesis tool optimises away logical qubit interconnects in the single logical qubit case. However, the overall utilisation of flip-flops and look-up tables in the design is very low ($<\SI{1}{\percent}$ of device resources), because the calculations involved in the design are quite simple. 

The use of I/O pads is quite high, due to the need for one measurement input $m$, one adaptive measurement setting $s$ and two byproduct operator lines per logical qubit. In our design, the total number of I/O pads required is
\begin{equation*}
  K = 4N + 4,
\end{equation*}
where $N$ is the number of logical qubits. This includes four common signals: the input clock $X_p$; the clock-is-locked output signal from the MMCM; a reset signal; and an enable signal. By accessing the byproduct operators via a low speed serial interface, it would be possible to reduce this pin count to
\begin{equation*}
  K \sim 2N,
\end{equation*}
which includes only the measurement inputs $m$ and adaptive measurement setting outputs $s$. On the largest FPGA in the 7-series family \cite{ds180}, the Virtex-7 xc7v2000t device (which has 1200 user I/O pads), this provides an upper bound on the number of logical qubits (cluster state rows) of $N\sim 600$. 

\begin{table}[t]
  \centering
  \begin{tabularx}{\columnwidth}{cyyyyyyyyyyc}
    \toprule
    & \multicolumn{3}{c}{Flip-flops} & \multicolumn{3}{c}{Look-up tables} &\multicolumn{3}{c}{Input/output} \\
    \cmidrule(lr){2-4}\cmidrule(lr){5-7}\cmidrule(l){8-9}
    N & CS & Full & Util. & CS & Full & Util. & Full & Util.\\
    \midrule 
    1 & 10 & 24 & \SI{0.03}{\percent} & 5 & 11 & \SI{0.03}{\percent} & 8 & \SI{2.8}{\percent}\\
    20 & 237 & 476 & \SI{0.89}{\percent} & 137 & 364 & \SI{0.58}{\percent} & 84 & \SI{29.5}{\percent}\\
    \bottomrule
  \end{tabularx}
  \caption{Utilisation of flip-flops, look-up-tables and input/output pads (I/O) in the design, for N=1 logical qubit and N=20 logical qubits, for the control system (CS) in Figure~\ref{fig:system} and the full design. The proportion of device resources is included in the utililisation (Util.) columns.}\label{tab:util}
\end{table}

Input/output delays are also a bottleneck for performance in the FPGA design, as we show in Section \ref{sec:timing}. The Xilinx 7-series devices were chosen because they have a level-sensitive latch built into their input logic slice (LCDE) \cite{ug471}, which forms the first stage of the digital system.

A disadvantage of the design is that it is not possible to place the output $s$ in the output logic slice, because there is combinational logic between the final register and the output port \cite{ug471}. It is also not possible to place the byproduct operator registers in output logic slices, because the output is rerouted to the internal FPGA fabric for use in updating the byproduct operators (see the feedback loop in Figure~\ref{fig:system}b).

As we show in Section \ref{sec:timing}, the clock frequency is not a bottleneck in the system, so it may be possible to create another design with multi-cycle latency, where the outputs are stored in separate registers and eligible for placing in the output logic slice. This may remove some of the output delay and allow a slightly higher clock frequency. It would also remove the logic hazards present in the output $s$.

\section{Verification of the design}
\label{sec:verification}

Due to the non-intuitive nature of the measurement patterns and the complexity of the digital hardware design, it is not possible to verify the functional correctness of the design simply by looking at the output of simulations. This section describes the verification of the measurement patterns and the program logic, and also the hardware design. 

\subsection{Measurement-based quantum computing simulator}

We wrote an MBQC simulator in C++ for the purpose of generating data to verify the digital system design. The program can simulate a cluster state containing up to 14 logical qubits by only holding two columns of the cluster state in memory at any one time.

The program is designed to mimic the operation of the hardware, using the program word $P$ to process measurement outcomes and apply quantum operations to the simulated quantum state according to the resulting adaptive measurement settings. At the end of the quantum circuit, the byproduct operators are applied to the state to obtain the result from the quantum computation.

The quantum circuit is also performed in the gate-based model on a state vector containing the same number of logical qubits. At the end, this state vector is compared with that obtained from the cluster state computation, to check that they agree with each other.

The state of the cluster state simulator at each measurement round is written to a file, which is used as the input to the hardware simulator. It contains the program word and the measurement outcomes, which are the inputs to the digital system. It also contains the value of the adaptive measurement settings, the byproduct operators and the stored byproduct operators, which are the outputs from the digital system.

\subsection{VHDL testbenches}

\Figure[t!](topskip=0pt, botskip=0pt, midskip=0pt)[width=0.99\textwidth]{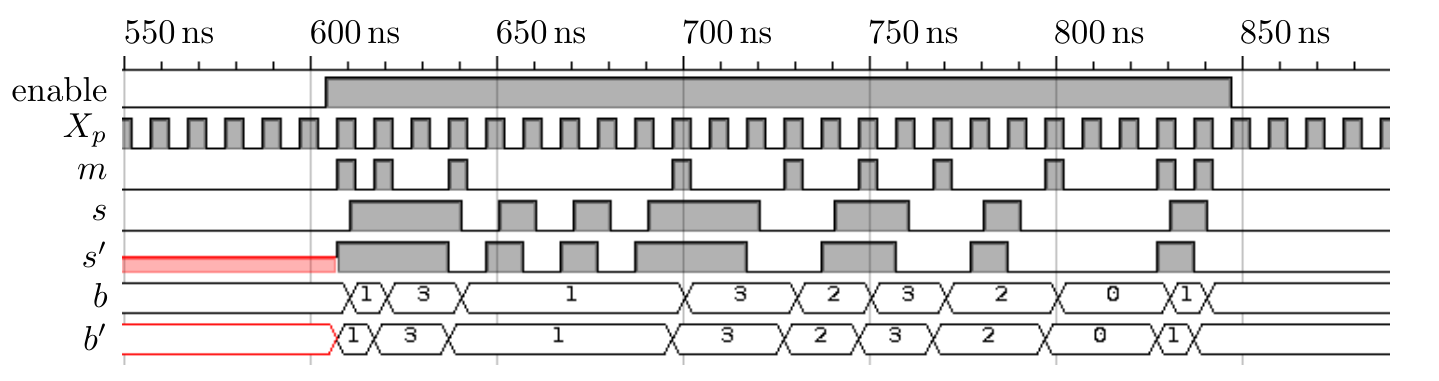}{An example of the post-implementation functional simulation. The outputs $s$ and $b$ are compared with the true values $s^\prime$ and $b^\prime$ from the simulation file. The hardware outputs slightly lag the true values because the file is loaded on $X_p$ in the testbench, whereas the design outputs the measurement settings and byproduct operators on the rising edge of $X_s$. \label{fig:sim}}

The function of the digital system was verified using testbenches written in VHDL. The testbenches read stimulus and output data from the simulation output file described in the previous section.

The output from the system, the adaptive measurement setting and the byproduct operators, are compared with the values from the simulation file. The simulation passes if all the values are equal, which is tested automatically. An example waveform output from the testbench, for a single logical qubit, is shown in Figure \ref{fig:sim}.

\section{Timing analysis}
\label{sec:timing}

We used static timing analysis to establish the maximum operating frequency of the design and to obtain the input/output delays associated with the system. The critical path is made up of two components:

\begin{itemize}
\item The path from the input port $m$ (clocked on the rising edge of $X_p$) to the byproduct operator register (loaded on the rising edge of $X_s$)
\item The path from the shift register output (loaded on the rising edge of $X_s$) to the output port $s$ (clocked on the rising edge of $X_p$).
\end{itemize}

By modifying the phase shift of $X_s$ relative to $X_p$, it is possible to allocate more time to one path or the other. The phase of $X_r$ must also be adjusted to allow timing closure of paths between the $X_s$ and $X_r$ clock domains. We established the maximum operating frequency $F_\text{max}$ of the system by manually adjusting the phase of $X_s$ and $X_r$ to balance the worst negative setup slack between the critical paths, while increasing the frequency of the design, until both paths fail to meet timing. Using this method, we obtained ${F_\text{max} = \SI{190}{\mega\hertz}}$ using $\text{phase}(X_s) = 220^{\circ}$ and $\text{phase}(X_r) = 300^{\circ}$. The phase difference \SI{80}{\degree} between $X_s$ and $X_r$ represents the amount of the time taken for the internal FPGA logic to process the latched measurement outcome before it is reset.

We then performed the timing analysis at each frequency between \SI{10}{\mega\hertz} and \SI{190}{\mega\hertz}, in steps of \SI{10}{\mega\hertz}, to establish the most generous input and output constraints that still allow timing closure at each frequency. All input/output constraints are expressed with respect to the external clock $X_p$ (the system clock).

The input constraint is specified by the clock-to-out time $t_{co}$ of the input signal $m$, which is equal to the time delay between the rising edge of $X_p$ and the pulse generated by the input analog system at $m$. This time constrains the analog characteristics of the single-photon detector amplifier.

The output constraint is the setup time $t_{su}$ of the output signal $s$ with respect to the system clock $X_p$, which is the delay between the time that $s$ transitions at the boundary of the FPGA and the next rising edge of $X_p$. This time determines the required operating speed of the output DAC system and modulator drivers, which must be able to set the voltages of the modulators before the next photon arrives on the rising edge of $X_p$.

The input/output timing constraints are plotted as a function of frequency in Figure \ref{fig:plot-inout}. The input constraint is systematically more generous than the output constraint, because of the choice of phase of $X_s$. The sum of the input and output constraints must be less than the total input/output slack, also shown in the figure.

\Figure[t!](topskip=0pt, botskip=0pt, midskip=0pt)[width=0.99\columnwidth]{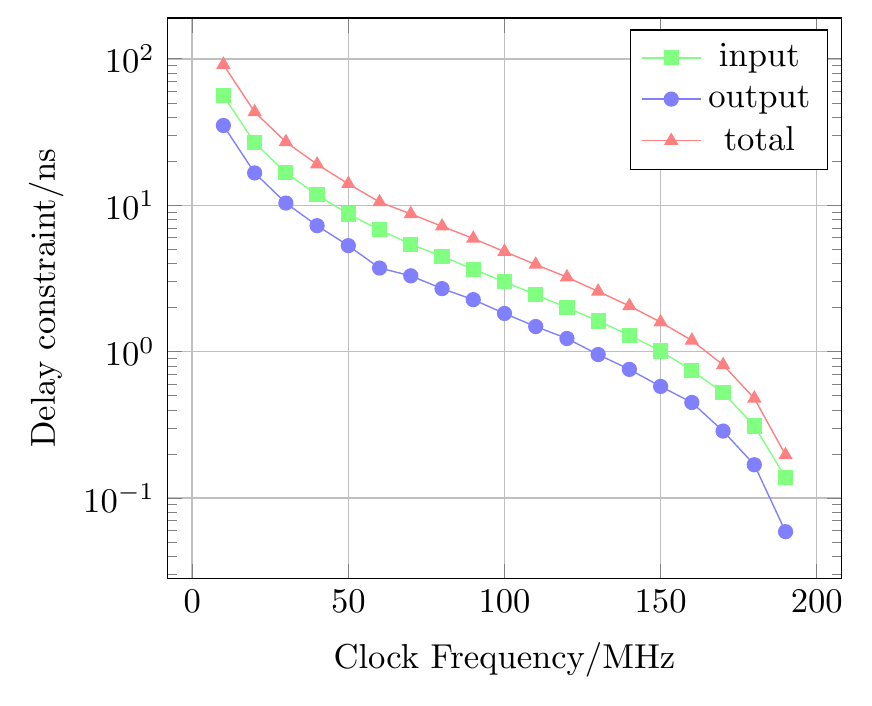}{The most generous input and output delay constraints that allow implementation of the design at each frequency. The total delay, which can be apportioned between input and output analog systems by adjusting the phase of $X_s$, represents the maximum amount of time available to the analog system shown in Figure \ref{fig:analog-digital}.  \label{fig:plot-inout}}

\Figure[t!](topskip=0pt, botskip=0pt, midskip=0pt)[width=0.99\columnwidth]{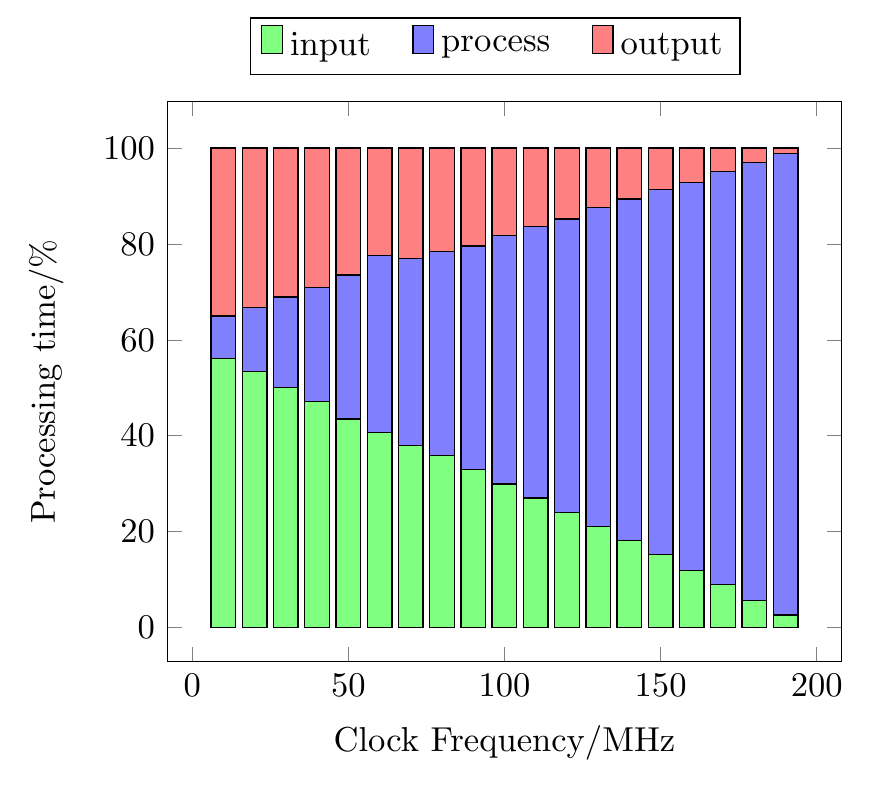}{The proportion of the clock cycle devoted to processing the adaptive measurement settings and the byproduct operators, as a function of photon clock frequency. At the higher frequencies, nearly all of the cycle is spent processing the measurements, leaving almost no time for the analog amplification at the input and output (shown in green and red). \label{fig:plot-proportion}}

Figure \ref{fig:plot-proportion} shows a graph of the proportion of the clock cycle $X_p$ taken up with digital processing, as a function of frequency. It is clear that at higher frequencies, the digital processing dominates the clock cycle, leaving very little time for the analog amplifier systems.

At a representative clock frequency of \SI{150}{\mega\hertz}, the photons would need to be delayed for \SI{6.67}{\nano\second} in either an optical fibre or a waveguide delay line. Assuming a standard silicon-on-insulator (SOI) platform, the delay line must be approximately \SI{83}{\centi\meter}, assuming a mode index of $\sim$ 2.4 \cite{chrostowski2015silicon}.

We re-implemented the design targeting a higher end FPGA (Xilinx Kintex Ultrascale+, part no. xcku5p-ffvd900-3-e), to see whether the maximum clock frequency could be improved. We found that the maximum clock frequency increased to ${F_\text{max} = \SI{220}{\mega\hertz}}$ using $\text{phase}(X_s) = 140^{\circ}$ and $\text{phase}(X_r) = 230^{\circ}$. In this case, at the maximum clock frequency, less time is allocated to the input analog system compared to the 7-series FPGA. The phase difference of \SI{90}{\degree} between $X_s$ and $X_r$ indicates that approximately the same time (\SI{1.125}{\nano\second}) is taken by the internal digital system compared to the 7-series FPGA (\SI{1.152}{\nano\second}).

\section{Extending the design to more realistic systems}
\label{sec:extensions}

The model of photonic quantum computing described here, with a deterministic cluster state generator, is a substantial simplification compared to what is required for a real photonic quantum computer. Firstly, we have ignored the question of cutting out the correct shaped cluster state for the given circuit. This would entail two rounds of measurement per column, because removing qubits causes measurement setting dependencies within a single column, as described in Section \ref{sec:cut-out}. Such dependencies will also be introduced by any measurement pattern with measurement settings that depend on the cluster qubits above and below.

For measurement patterns like the CPhase gate \cite[Section IV.C.]{Raussendorf2003}, it is necessary to be able to arbitrarily re-order the measurements between pairs of columns. This is to satisfy the requirement that measurement outcomes are always available before any dependent adaptive measurement settings are required. Depending on the complexity of the measurement pattern, it may be necessary to merge more than two measurement columns. The corresponding system for arbitrarily rerouting photons into different measurement blocks and processing the outcomes would grow in complexity as a result.

Another complexity that may arise out of more complicated measurement patterns is the increased non-locality of the byproduct operator calculation. In the design discussed in this paper, the byproduct operators depend only on measurement outcomes from adjacent logical qubits. However, there are measurement patterns for which byproduct operators for a given logical qubit may depend on cluster qubits that are further away \cite[Section IV.C.]{Raussendorf2003}. This may lead to a routing problem in FPGA and ASIC designs, especially as the number of qubits increases.

An important complication is the lack of a deterministic cluster state generator. There are schemes for generating probabilistic cluster states (i.e. cluster states that may contain missing edges); this non-determinism must be overcome by tracking the successfully generated cluster state edges in real time, and mapping the measurement pattern dynamically onto the resulting graph \cite{Kieling2007}.

In this scheme, it would be the necessary to hold many columns of the cluster state in delay lines at a time, rather than just one, in order to keep enough depth to correct for dead-ends in the partial cluster state. For a given delay line length, the photon clock frequency $X_p$ would have to be proportionally increased in order to fit multiple photons in the same length of delay line. At the same time, the algorithmic complexity of the digital system would increase substantially with the inclusion of a real-time algorithm to track the structure of the unfolding cluster state before the adaptive measurement calculation.

On the other hand, it is possible to optimise the measurement patterns for ease of implementation in a photonic setting. For example, by inserting the identity pattern ($I$ in Figure~\ref{fig:cluster-state}b) between the $X$ and $Z$ rotations in the arbitrary one-qubit gate~\cite{2101.09310}, the timing requirements on the adaptive measurement setting could be significantly reduced, at the expense of longer measurement patterns. This approach is related to the percolation system described above, where logical qubit ``wires'' can be implemented using the identity pattern and the control system is responsible for placing measurement patterns on cluster qubits in real time.

Finally, if an SNSPD is used as the single-photon detector, the electronic control system may be required to operate at cryogenic temperatures so as to avoid data delays into and out of the cryostat. In this case, static timing analysis using a program such as Vivado is not valid. A 7-series FPGA has been found to operate at \SI{4}{\kelvin}, with a slight performance increase across many metrics \cite{Charbon2021}. However, timing jitter inside the FPGA increases slightly, meaning that timing closure may not be achievable close to $F_\text{max}$.

The complex set of factors described above in a realistic photonic MBQC design makes it impossible to make general statements about the performance of classical control systems in this setting. In order to establish quantitative performance bounds, it is necessary to implement and analyse a simple example system incorporating these more realistic details.

\section{Discussion}
\label{sec:discussion}

In contrast to every other approach to building quantum computers, photonic MBQC relies on manipulating and measuring flying-qubit states. This means that the effective `lifetime' of a qubit in these platforms is ultimately bounded by the length of time that photons can be kept circulating inside an optical delay line, either on or off-chip. The fact that the spatial and temporal properties of the system can not be decoupled is at the root of many of the unique timing constraints that photonic approaches need to satisfy. This is in contrast to other matter-based systems where the qubit lifetime is, to first order, unrelated to its spatial footprint.

In an integrated photonic approach, the only way to get longer qubit lifetimes is by increasing the length of the on-chip delay line.  Even with a high-index-contrast platform like SOI, which allows low-loss bend radii $<\SI{5}{\micro\meter}$, getting realistic delays beyond \SIrange{2}{3}{\nano\second} is extremely challenging, both due to the increasing insertion loss (\SIrange{1}{2}{\decibel\per\centi\meter}) and the increasing on-chip footprint (spiral delay lines with lengths $\sim$ 10 cm) \cite{chrostowski2015silicon}. One solution to the timing constraints is to use an integrated photonic quantum memory \cite{Ma2020} which would make photonic MBQC implementations closer to their matter-based counterparts by allowing one to map quantum information on to a long-lived spin / hyperfine transition.

Longer delays can be obtained in principle by using low-loss optical fibers off-chip, although this approach is not without its own trade-offs. Losses in the grating couplers involved in getting the light on and off the chip must be accounted for, in addition to losses involved in the optical switching network needed to get the cluster states to the grating couplers. These requirements can in principle be satisfied by state-of-the-art lithium niobate modulators, however the size and form factor are not really suitable for very large scale integration, which is a critical requirement from a systems perspective. State-of-the-art silicon modulators are very far from ideal, especially in terms of insertion loss ($\sim \SI{6}{\decibel}$ / device) \cite{Witzens2018}.

It is likely that a performance improvement could be obtained by implementing the digital design using an ASIC. Critical path delays due to logic have been found to decrease by 3-4 times in standard-cell ASIC designs~\cite{Kuon2007}. However, this may not translate to a performance improvement in this design because the majority of the critical path delays come from the input/output buffers, not the logic. To improve this, it may be possible to utilise very high speed latches and output buffer designs, with delays on the order of \SI{100}{\pico\second} \cite{Heydari}. A full analysis of the input/output buffer delays should be performed in tandem with the design of the input/output analog systems, to ensure compatibility between the two systems. At this point, the requirement for absolute synchronisation between the cluster state generator and the digital control system, using a system synchronous architecture \cite{athavale2021high} may become the bottleneck to the design. Such schemes are often limited to speeds up to \SIrange{200}{300}{\mega\hertz}, due to clock skew and data path delays \cite{hall2000high}.

\section{Conclusion}
\label{sec:conclusion}

We have provided a practical description of the measurement patterns for one-qubit gates and the CNOT gate and shown in detail how to implement a digital control system for photonic MBQC, in the presence of an ideal path-encoded photonic cluster state generator. It is clear from the timing analysis of our FPGA implementation of this system that it places substantial constraints on the input and output analog systems needed at the interface between the classical and quantum subsystems. For example, at a photon clock frequency of \SI{150}{\mega\hertz}, the total time available for the input and output analog processing is \SI{1.59}{\nano\second} out of the total period \SI{6.67}{\nano\second}. The remaining \SI{5.08}{\nano\second} is consumed by the logic delays inside the FPGA design. At the same time, a photon clock period of \SI{6.67}{\nano\second} corresponds to a long delay line ($\sim$ 83 cm), that will occupy quite a large footprint in an integrated implementation of photonic MBQC.

While in this work we have implemented a proof-of-principle design to study the constraints, it is clear that the digital system and implementation can be further optimised. For example, since the maximum frequency of our design is less than \SI{200}{\mega\hertz} and the maximum clock frequency of the target FPGA is greater than \SI{600}{\mega\hertz}, it may be possible to create a multi-cycle digital design so as to properly register the inputs and outputs and place them in dedicated input/output slices. This would likely increase the maximum clock frequency somewhat while maintaining the input/output delay constraints.

Incorporating the features of a realistic photonic MBQC system, where the cluster state generator is probabilistic, adds an additional level of algorithmic complexity to the design. It is likely that analysis and implementation of a minimal system design is necessary to address the additional overheads that are involved in these settings.

We would like to emphasize that most of the constraints considered in this paper are practical, rather than fundamental in nature. We have shown that, for our reference design, the input and output analog systems must have a combined latency on the order of hundreds of picoseconds. While these specifications might be challenging, they also provide an exciting design opportunity for classical high-speed optoelectronics. In our view, the viability of photonic MBQC rests on the improvement of the classical subsystem just as much as it does on improving the quantum resources (cluster state generation). Understanding how these timing constraints can be relaxed to the point where they can be satisfied by existing circuit architectures/amplifier topologies is a necessary near term-goal. We believe this can be achieved by analysing ASIC implementations of the circuits discussed here, with a focus on reducing the input/output routing and buffer delays which limit the maximum operating frequency of the design. On the other hand, there is also a need to rethink the photonic MBQC architecture with latency as the primary constraint and explore the improvements that can be achieved through architectural changes in the design.

\appendices

\section{Gate based quantum computing}
\label{app:gate-based-qc}

This appendix contains a brief overview of quantum computing in the gate-based model. The basic unit of quantum computation is the qubit, which is a two-state system, analogous to a bit, except complex linear combinations of the zero-state (denoted $|0\rangle$) and the one-state (denoted $|1\rangle$) are also valid states. The states $|\psi\rangle$ of a qubit can be expressed as
\begin{equation}
  \label{eq:qubit-state}
  |\psi\rangle = a|0\rangle + b|1\rangle,\quad a\in\mathbb{R},b\in\mathbb{C}.
\end{equation}

The qubit can only ever be observed in the state $|0\rangle$ or $|1\rangle$, with probabilities given by the ratio of $|a|^2$ to $|b|^2$. The act of observing the qubit is called a measurement. The absolute values of $a$ and $b$ have no independent physical meaning, so the condition $|a|^2 + |b|^2=1$ is imposed so that the probabilities are equal to $|a|^2$ and $|b|^2$. Likewise, the arguments of the complex numbers $a$ and $b$ have no physical meaning, so it is possible to impose $a\in\mathbb{R}$ without loss of generality. The argument of $b$ is then the relative phase between $|0\rangle$ and $|1\rangle$.

The states of a single qubit can be identified with points on the surface of a sphere, called the Bloch sphere, as shown in Figure \ref{fig:bloch-sphere}. The mapping between the $a$ and $b$ and the real number angles $\theta$ and $\phi$ is given by the identity:

\[
  a|0\rangle + b|1\rangle = \cos(\theta/2)|0\rangle + e^{i\phi}\sin(\theta/2)|1\rangle.
\]
The angle $\phi$ in the equator of the Bloch sphere is the relative phase between $|0\rangle$ and $|1\rangle$, and the angle $\theta$ controls the probability of observing $|0\rangle$ or $|1\rangle$ upon measurement.

\subsubsection{One-qubit gates}

The state of the qubit can be changed by applying a quantum gate. The valid gates on a single qubit, called one-qubit gates, are those which correspond to a rotation of the points on the Bloch sphere about any axis, by any angle. The gates which perform rotations of the state about the $x$, $y$ and $z$ axes are denoted $R_x(\alpha)$, $R_y(\alpha)$ and $R_z(\alpha)$, where $\alpha$ is the angle of rotation according to the right-hand rule. An arbitrary one-qubit rotation can be formed by applying $x$- and $z$-rotations in sequence as $R_x(\zeta)R_z(\eta)R_x(\xi)$ (applied from right to left). This follows from the decomposition using Euler angles of an arbitrary rotation into $x$- and $z$-rotations.

\subsubsection{Measurement}

When a qubit is measured, it always collapses to either the state $|0\rangle$, with probability $|a|^2$, or the state $|1\rangle$, with probability $|b|^2$. This is called a computational basis measurement.

However, it is possible to generalise the concept of measurement so that an ``observation'' causes the qubit to collapse into the state $|0^\prime\rangle$ or the state $|1^\prime\rangle$, which are any two antipodal points on the Bloch sphere, joined by a line $L$. This observation is made by using one-qubit gates to transform the line $L$ to the line through $|0\rangle$ and $|1\rangle$, and then making a computational basis measurement. For example, to measure along the line denoted $L$ in Figure \ref{fig:bloch-sphere}, it is necessary to apply a $z$-rotation $R_z(-\phi+\pi/2)$ to align the state $|0^\prime\rangle$ with the positive $y$ axis, followed by an $x$-rotation $R_x(\pi/2)$ to obtain $|0\rangle$.

It is possible to measure along any line in this way by applying an arbitrary one-qubit gate $R_z(\alpha)R_x(\beta)R_z(\gamma)$ and then measuring in the computational basis. It is important to realise that general measurements involve the application of a one-qubit gate before making a computational basis measurement. 

\subsubsection{Two-qubit gates}

The states of two qubits can be expressed analogously to Equation (\ref{eq:qubit-state}) as
\begin{equation}
  \label{eq:two-qubit-state}
  |\psi\rangle = a|00\rangle + b|01\rangle + c|10\rangle + d|11\rangle,
\end{equation}
where $a\in\mathbb{R}$ and $b,c,d\in\mathbb{C}$. The sum is over all the four possible states that the two qubits could be observed in. As with the single qubit case, $|a|^2+|b|^2+|c|^2+|d|^2=1$ is imposed, and the probability of obtaining, for example,  $|01\rangle$, is given by $|b|^2$.

There is no equivalent of the Bloch sphere for graphically presenting the states of two qubits. An example of a two-qubit gate is the CNOT gate. The action of this gate on the state (\ref{eq:two-qubit-state}) above is
\begin{multline}
  |\psi\rangle = a|00\rangle + b|01\rangle + c|10\rangle + d|11\rangle \\\mapsto   |\psi\rangle = a|00\rangle + b|01\rangle + c|11\rangle + d|10\rangle,
\end{multline}
that is, the states $|10\rangle$ and $|11\rangle$ are reversed. The interpretation of this gate is that the first (leftmost) qubit controls whether a NOT gate is applied to the second (rightmost) qubit. The first qubit is called the control qubit, and the second qubit the target. 

Analogously to the way that a NAND gate is universal for digital logic, the CNOT gate combined with the basic rotations $R_x(\alpha)$, $R_y(\alpha)$ and $R_z(\alpha)$ are universal for quantum computation. To build up any complicated computation, all that is required is to apply the correct string of one- and two-qubit gates, one after the other, to a set of qubits. For example, in Figure~\ref{fig:cluster-state}c, an arbitrary one-qubit gate $U=R_x(\zeta)R_z(\eta)R_x(\xi)$ is applied to the top qubit, and a CNOT gate is applied between the bottom two qubits.

\section{CNOT Measurement Pattern}
\label{app:cnot-derivation}

\Figure[t!](topskip=0pt, botskip=0pt, midskip=0pt)[width=0.99\columnwidth]{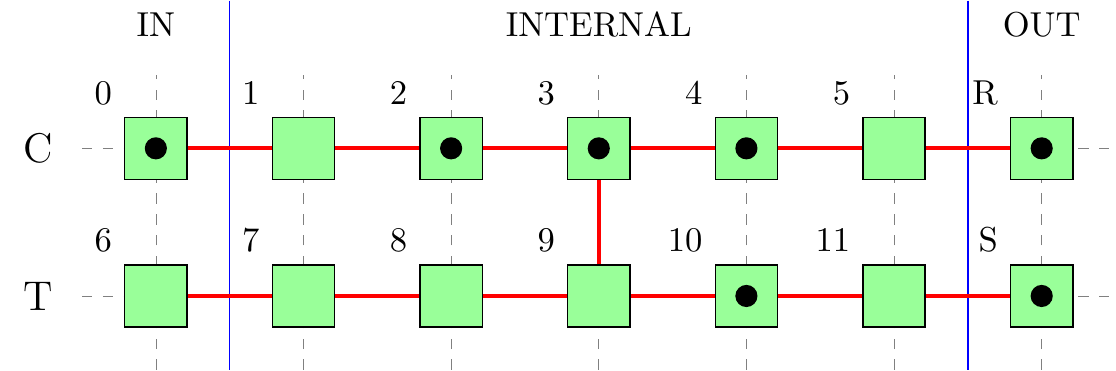}{The labelling of the cluster qubits for the purpose of deriving the CNOT measurement pattern. When a gate is realised in MBQC, the input state starts on the IN column and is teleported to the OUT column $R$ and $S$ by applying the measurement pattern. The black dots show the location of the correlation operators $K_a$ in Equation (\ref{eq:correl}) below.\label{fig:cnot-derive}} 

We use a reduced measurement pattern for the CNOT gate that only uses two rows of cluster qubits, instead of the three row pattern in \cite{Raussendorf2003}. The pattern is derived using the same method outlined in Section II.G.7. of that paper for the calculation of the three-row CNOT gate. In order to explain the derivation, we begin by discussing some technical aspects of cluster states, and describe what it means for a measurement pattern to realise a gate.

A cluster state $|\phi_C\rangle$ on $N$ qubits is created by placing all the qubits in the $|+\rangle$ state, and then applying CZ gates between each pair of qubits that should have an entanglement link (shown as red line segments in Figure~\ref{fig:cnot-derive}). It can be shown \cite{Raussendorf2003} that cluster states satisfy the eigenvalue equations
\begin{equation}
  \label{eq:correlation-ops}
K_a|\phi_C\rangle = \left(X_a \prod_{b\sim a} Z_b\right)|\phi_C\rangle = |\phi_C\rangle,
\end{equation}
where the first equality defines the correlation operator $K_a$ on the cluster qubit $a$. There is one such equation for each cluster qubit $a$, and in each equation, the product is over all other neighbouring cluster qubits $b$ joined by red line segments to $a$ (denoted $b\sim a$).

To state what it means for a measurement pattern to realise a gate $G$, we use to the arrangement of qubits shown in Figure~\ref{fig:cnot-derive}, on which the CNOT measurement pattern is defined. Instead of placing all the qubits in the $|+\rangle$ state, assume qubits 0 and 6 (the IN qubits) are in an arbitrary state $|\phi\rangle$. As before, place all the other qubits (including the OUT qubits) in the $|+\rangle$ state, and apply CZ gates wherever there are red line segments in the Figure~\ref{fig:cnot-derive}. Now, after the measurement pattern for the CNOT gate has been applied, meaning that all the IN and INTERNAL qubits have been measured out, there remains a two-qubit state $|\psi\rangle$ on the OUT qubits $R$ and $S$. The sense in which the measurement pattern has realised the gate $G$ is that input and output states are related by 
\begin{equation}
  |\phi\rangle = BG|\psi\rangle,
\end{equation}
where $B$ is the byproduct operator for the measurement pattern. In other words, the measurement pattern has the effect of moving the state of the IN column to the OUT column, and transforming it according to the gate which is being realised by the measurement pattern.

The measurement pattern for the CNOT gate is obtained by using a theorem \cite[Theorem 1]{Raussendorf2003} that relates eigenvalue equations derived from Equation (\ref{eq:correlation-ops}) and a given measurement pattern, to the gate $G$ which that measurment pattern realises. The content of the theorem is that it is only necessary to check how a cluster state $|\phi_C\rangle$ is affected by the measurement pattern (where the state of qubits 0 and 6 are $|+\rangle$) in order to establish that the measurement pattern works for any other IN state $|\phi\rangle$. In the interest of simplicity, We state the theorem for the case of a two-qubit gate $G$ like the CNOT gate:

\begin{thm}
  Suppose that a cluster state $|\phi_C\rangle$ is prepared on the pattern of 14 qubits shown in Figure~\ref{fig:cnot-derive}, for the purpose of realising a two-qubit gate $G$ acting on logical qubits labelled $C$ and $T$. Suppose that a set of measurements $M$ is performed on the INTERNAL cluster qubits 1 to 5 and 7 to 11, resulting in a state $|\psi_C\rangle$ on the remaining qubits (0, 6, $R$ and $S$), which satisfies the following sets of eigenvalue equations:
  \begin{equation}
    \label{eq:thm-1}
    \begin{aligned}
      X_0\left[GX_CG^\dagger\right]_{R,S}|\psi_C\rangle &= (-1)^{\lambda_x}|\psi_C\rangle\\
      Z_0\left[GZ_CG^\dagger\right]_{R,S}|\psi_C\rangle &= (-1)^{\lambda_z}|\psi_C\rangle
    \end{aligned}
  \end{equation}
  and
  \begin{equation}
    \label{eq:thm-2}
    \begin{aligned}
      X_6\left[GX_TG^\dagger\right]_{R,S}|\psi_C\rangle &= (-1)^{\mu_x}|\psi_C\rangle\\
      Z_6\left[GZ_TG^\dagger\right]_{R,S}|\psi_C\rangle &= (-1)^{\mu_z}|\psi_C\rangle
    \end{aligned}
  \end{equation}
  Then the measurement pattern in which the inner qubits are measured according to $M$, and the IN cluster qubits 0 and 6 are measured in the $X$-basis, realises the gate $GB$, where the byproduct operators $B$ for the logical qubits $C$ and $T$ are given by
  \begin{equation}
    \label{eq:thm-byproduct}
    \begin{aligned}
      (x_C, z_C) &= (\lambda_z, m_0 + \lambda_x)\\
      (x_T, z_T) &= (\mu_z, m_6 + \mu_x),
    \end{aligned}
  \end{equation}
  where $m_a$ is the outcome of the measurement of the $a^\text{th}$ cluster qubit.\qed
\end{thm}

The square bracketed terms in Equations (\ref{eq:thm-1}) and (\ref{eq:thm-2}) are computed in terms of the logical qubits $C$ and $T$, without reference to cluster qubits. Any terms involving $C$ and $T$ are then interpreted as applying to the cluster qubits $R$ and $S$. For example, when $G=\text{CNOT}$, $${\left[GX_TG^\dagger\right]_{R,S} = [X_CX_T]_{R,S}} = X_RX_S.$$ To apply the theorem to the CNOT gate, it is therefore necessary to obtain the following eigenvalue equations
  \begin{equation}
    \label{eq:cnot-1}
    \begin{aligned}
      X_0\left(X_RX_S\right)|\psi_C\rangle &= (-1)^{\lambda_x}|\psi_C\rangle\\
      Z_0\left(Z_R\right)|\psi_C\rangle &= (-1)^{\lambda_z}|\psi_C\rangle
    \end{aligned}
  \end{equation}
  and
  \begin{equation}
    \label{eq:cnot-2}
    \begin{aligned}
      X_6\left(X_S\right)|\psi_C\rangle &= (-1)^{\mu_x}|\psi_C\rangle\\
      Z_6\left(Z_RZ_S\right)|\psi_C\rangle &= (-1)^{\mu_z}|\psi_C\rangle
    \end{aligned}
  \end{equation}

To obtain these equations, begin with the cluster state $|\phi_C\rangle$ on the two-row CNOT shape shown in Figure \ref{fig:cnot-derive}, and multiply together the correlation operators in Equation (\ref{eq:correlation-ops}) so as to obtain the following four equations:
\begin{align}
  \label{eq:correl}
  |\phi_C\rangle &= K_0K_2K_3K_4K_RK_{10}K_S|\phi_C\rangle\\\nonumber
&\qquad=-\textcolor{red}{X_0}Y_2X_3\textcolor{blue}{Y_4}\textcolor{red}{X_R}X_{10}\textcolor{red}{X_S}|\phi_C\rangle\\\nonumber 
  |\phi_C\rangle &= K_1K_2K_4K_5|\phi_C\rangle \\\nonumber
                 &\qquad=\textcolor{red}{Z_0}Y_1Y_2\textcolor{blue}{Y_4}Y_5\textcolor{red}{Z_R}|\phi_C\rangle\\\nonumber
  |\phi_C\rangle &= K_6K_8K_{10}K_S|\phi_C\rangle\\\nonumber &\qquad=\textcolor{red}{X_6}X_8X_{10}\textcolor{red}{X_S}|\phi_C\rangle\\\nonumber
  |\phi_C\rangle &= K_4K_5K_7K_9K_{11}|\phi_C\rangle\\ &\qquad =\textcolor{blue}{Y_4}Y_5\textcolor{red}{Z_R}\textcolor{red}{Z_6}X_7X_9X_{11}\textcolor{red}{Z_S}|\phi_C\rangle.\nonumber
\end{align}

The right hand sides are obtained by repeated application of the equation $X_aZ_a=iY_a=-Z_aX_a$. Note that Pauli operators on different qubits commute.

As with any pattern derived using this method, the choice of operators $K_a$ in the above equations is motivated by two goals
\begin{itemize}
\item The equations must contain the correct IN and OUT terms in Equations (\ref{eq:cnot-1}) and (\ref{eq:cnot-2}). These terms are coloured red in the equations;
  \item The Pauli operators on the INTERNAL cluster qubits agree between all the equations. That is, for each cluster qubit $a$, only $X_a$ or $Y_a$ appears across all the equations. For example, when $a=4$, only $Y_4$ appears (three times, shown in blue), and there are no instances of $X_4$. It is these operators that define the measurement bases $M$ for each qubit $a$ in the INTERNAL group of cluster qubits.
\end{itemize}

When the INTERNAL qubits are measured according to $M$, the Pauli terms disappear \cite[Section 10.5.3]{Nielsen2009}, and each one contributes a sign according to its measurement outcome $m_a$, to give the following equations on the reduced state $|\psi_C\rangle$:
\begin{align*}
X_0X_RX_S|\psi_C\rangle &= (-1)^{1+m_2+m_3+m_4+m_{10}}|\psi_C\rangle\\
Z_0Z_R|\psi_C\rangle &= (-1)^{m_1+m_2+m_4+m_5}|\psi_C\rangle\\
X_6X_S|\psi_C\rangle &= (-1)^{m_8+m_{10}}|\psi_C\rangle\\
Z_6Z_RZ_S|\psi_C\rangle &= (-1)^{m_4+m_5+m_7+m_9+m_{11}}|\psi_C\rangle
\end{align*}
These equations are in the form of Equations (\ref{eq:cnot-1}) and (\ref{eq:cnot-2}), and define the values of $\lambda_x,\lambda_x,\mu_x,\mu_z$ in terms of the measurement outcomes $m_a$. As a result, it follows from the theorem above that the measurement pattern consisting of $M$, plus $X$ measurements on the IN qubits, realises the gate $(\text{CNOT})B$, where the byproduct operator $B$ found using Equation (\ref{eq:thm-byproduct}) to be
\begin{equation}
  \begin{aligned}
    (x_C, z_C) = (&m_1+m_2+m_4+m_5, \\
    &1+m_0+m_2+m_3+m_4+m_{10})\\
    (x_T, z_T) = (&m_4+m_5+m_7+m_9+m_{11},\\
    &m_6 + m_8+m_{10}).
  \end{aligned}
\end{equation}

Finally, the byproduct operator can be commuted past the CNOT gate to obtain
\begin{multline}
(Z_C^{1+m_0+m_2+m_3+m_4+m_6+m_8} X_C^{m_1+m_2+m_4+m_5}\\Z_T^{m_6 + m_8 + m_{10}} X_T^{m_1+m_2+m_7+m_9+m_{11}})\text{CNOT}.
\end{multline}
The contributions to the byproduct operators given in this formula are depicted in Figure \ref{fig:basic-patterns}, and stated in Equations (\ref{eq:cnot-byp-1}) and (\ref{eq:cnot-byp-2}).

\section*{Acknowledgment}

JRS would like to thank Lana Mineh for help working out the reduced CNOT measurement pattern, and for assistance in programming the C++ MBQC simulator, and Oliver Thomas for many interesting discussions regarding the implementation of photonic quantum computing. We would like to thank Jose Nunez-Yanez for very helpful discussions regarding FPGA design. JRS received funding from the Bristol Quantum Engineering Center for Doctoral Training, EPSRC Grant No. EP/L015730/1.
KCB would like to thank the European Research Council for funding support (ERC-StG SBS3-5, 758843). 

%% Paper contents ends here %%%%%%%%%%%%%%%%%%%%%%%%%%%%%%%%%%%%%%%%%%%%%%%

\bibliographystyle{mybibstyle}
\bibliography{ref}

\end{document}